%% file: main.tex
\documentclass[preprint,12pt]{elsarticle}

\usepackage{upgreek}

\usepackage{color}
\usepackage{url}
\usepackage{fancyvrb}
\usepackage{eurosym}
\usepackage[title]{appendix}
\usepackage{amsmath}
\usepackage{breqn}

\usepackage{soul}
\newcommand{\rev}[1]{{#1}}

\usepackage{tikz}
\usetikzlibrary{arrows,shapes,positioning,shadows,trees}

\tikzset{
  basic/.style  = {draw, text width=7.5em, font=\sffamily, rectangle},
  root/.style   = {basic, rounded corners=2pt, thin, align=center,
                   fill=none, font=\sffamily},
  level 2/.style = {basic, rounded corners=6pt, thin,align=center, fill=white,
                   text width=7.5em, font=\small\sffamily},
  level 3/.style = {basic, thin, align=left, draw=none, fill=none, text width=7em, font=\footnotesize\sffamily}
}

\usepackage{graphicx}
\usepackage{xcolor}
\usepackage{ragged2e}
\usepackage{wrapfig}
\usepackage{lscape}
\usepackage{rotating}
\usepackage{epstopdf}
\usepackage{multicol}
\usepackage{amssymb}
\usepackage[modulo]{lineno}

\newcommand{\FogTorch}{\protect{{\sf\small FogTorch{\em $\Uppi$}}}}
\newcommand{\secfog}{\protect{{\sf\small SecFog}}}
\newcommand{\problog}{\protect{{\sf\small ProbLog}}}

\newcommand{\CR}{\protect{{\sf\small SR}}}
\newcommand{\ND}{\protect{{\sf\small ND}}}
\newcommand{\example}{\protect{{\noindent\bf Example.}}}
\newcommand{\asecfog}{\protect{{\sf\small $\alpha$SecFog}}}

\journal{Future Generation Computer Systems}

\begin{document}

\begin{frontmatter}


\title{Secure Cloud-Edge Deployments, with Trust}



\author{S. Forti\footnote{Corresponding author: {\tt\footnotesize stefano.forti@di.unipi.it}.\\ \\{\scriptsize \textcopyright 2019. This manuscript version is made available under the CC-BY-NC-ND 4.0 license \url{http://creativecommons.org/licenses/by-nc-nd/4.0/}. This manuscript is a preprint of the article published in FGCS and available at: \url{https://doi.org/10.1016/j.future.2019.08.020}} }, G.-L. Ferrari, A. Brogi}

\address{Department of Computer Science, University of Pisa, Italy}

\begin{abstract}
Assessing the security level of IoT applications to be deployed to heterogeneous Cloud-Edge infrastructures operated by different providers is a non-trivial task.
In this article, we present a methodology that permits to express security requirements for IoT applications, as well as infrastructure security capabilities, in a simple and declarative manner, and to automatically obtain an explainable assessment of the security level of the possible application deployments. The methodology also considers the impact of trust relations among different stakeholders using or managing Cloud-Edge infrastructures.
A lifelike example is used to showcase the prototyped implementation of the methodology.
\end{abstract}


\end{frontmatter}


\input{src/1_Intro.tex}
\input{src/_Background.tex}
\input{src/_Methodology.tex}

\input{src/_Example.tex}

\input{src/_asecfog.tex}

\input{src/_Related.tex}
\input{src/_Conclusions.tex}

\medskip
\noindent{\footnotesize \bf Acknowledgments}
{\footnotesize\noindent This work has been partly supported by the project ``{\it DECLWARE: Declarative methodologies of application design and deployment}" (PRA\_2018\_66) funded by the University of Pisa, Italy.
}


\bibliographystyle{model1-num-names}
\bibliography{sample}







\end{document}

%% file: src/1_Intro.tex
\section{Introduction}
\label{introduction}

\noindent
Enforcing Quality-of-Service (QoS) requirements in the deployment of Internet-of-Things (IoT) software systems is a non-trivial -- yet necessary -- task to accomplish. Indeed, such systems are often developed in large, highly distributed, multi-service architectures, which are to be deployed to complex, heterogeneous and highly distributed infrastructures, spanning the Cloud-IoT continuum. 
Cloud-Edge computing \cite{cloudedgecomputing} extends Cloud computing towards the edge of the Internet to better support latency-sensitive and bandwidth-hungry IoT applications by mapping service application functionalities (e.g., device management, telemetry ingestion, processing and storage, status and notification, multi-level analytics and data visualisation \cite{familiar2015iot}) wherever it is ``best-placed" to suitably meet all the application (hardware, software and QoS) requirements \cite{book, bonsai}. 

Various works (e.g., \cite{004,101,024, 011, 012, summersoc,howtoplace2019}) have tackled the problem of determining optimal placements (and management) of application services, mainly taking into account resource usage, deployment costs, network QoS (i.e., latency, response time, bandwidth), and energy consumption.
Considering these aspects all together is of primary importance in Cloud-Edge scenarios, where many life-critical (e.g., e-health, autonomous vehicles) or mission-critical (e.g., drone packet deliveries, smart farming) application verticals can significantly suffer from performance degradation due to bad resource allocation or insufficient Internet connectivity. Analogously, their management might aim at reducing operational costs, due to power consumption or to resource leasing, so to increase profit. 

However, to the best of our knowledge no approach has been proposed that accounts for the security requirements of the application to be deployed and matches them to the security capabilities available at different infrastructure nodes. Security of ICT solutions is an intrinsically complex problem, which requires reasoning about a system model by analysing its security properties and their effectiveness against potential attacks, and accounting for trust relations among different involved stakeholders (e.g., infrastructure and application operators). 
In addition, any methodology for optimal service placement that accounts for security, should in principle enable decision-makers to understand \textit{why} a certain deployment can be considered optimal, i.e. the provided recommendations should be \textit{explainable}. Indeed, explainable artificial intelligence (XAI) techniques are getting more attention from the security community since they can provide a concise explanation (\textit{proof}) of the query results \cite{vigano2018explainable}. In the case of determining secure and trustworthy deployments of an application, for instance,  XAI aims at answering questions like: \textit{Why is this deployment more secure than this other? Why and how much are they secure?}

As an extension to the Cloud, Cloud-Edge will share with it many security threats, while including its new peculiar ones. 
On the one hand, Cloud-Edge will increase the number of security enforcement points by allowing local processing of private data closer to the IoT sources. 
On the other hand, new infrastructures will have to face brand new threats for what concerns the physical vulnerability of devices. 
Indeed, application deployments to Cloud-Edge infrastructures will \rev{often} include accessible (Edge or IoT) devices that may be easily hacked, broken or even stolen by malicious users \rev{and that can only offer a limited set of security capabilities} \cite{ni2017securing, ambrosin2016sana}. 

Last but not least, as Cloud-Edge application deployments will likely span various service providers (some of which may be not trustworthy), trust must be considered when deciding where to place application services. \rev{Trust relations are especially important at the edge of the Internet, due to the uncertainty derived from dealing with multiple
infrastructure providers. Indeed, the security levels and reputation of edge capabilities might be as heterogeneous as their hardware capabilities, and might give rise to \textit{security} \textit{zones}, i.e. collections
of assets that share the exposure to certain security
risks} \cite{fmectrust}. Thus, in Cloud-Edge scenarios, where part of the application could be deployed to \rev{a mixture of well-established} (e.g., Clouds, ISPs) and opportunistic infrastructures (e.g., crowd-computing, \textit{ad-hoc} networks \cite{garcia}), \textit{trust models} are needed to estimate trust levels towards unknown providers, \rev{possibly} aggregating also trusted providers' opinions.

All this considered, the move of utility computing towards the edge of the network -- and in continuity with existing Clouds -- calls for new \textit{quantitative} and \textit{explainable methodologies} that permit to assess the security level of distributed multi-service IoT applications. Such methodologies should take into account application requirements, security countermeasures featured by the Cloud-Edge infrastructure, and trust relations in place among different stakeholders that manage or use Cloud-Edge infrastructures. 

\medskip
\noindent
In this paper, we propose a first step towards \rev{well-founded and declarative} reasoning methodologies \rev{for security- and trust-aware multi-service application deployment in Cloud-Edge scenarios}. Our proposal, \secfog, helps application operators in Cloud-Edge scenarios in determining the most secure application deployments by reducing manual tuning and by considering specific application requirements, infrastructure capabilities and trust.
\secfog\ has been prototyped in the \problog\ language \cite{problog07} and, as we will show, the prototype can be used together with existing approaches that solve the problem of mapping IoT application services to Cloud-Edge infrastructures according to requirements other than security and trust \rev{(e.g., hardware, network QoS, cost)}. 
\rev{A generalised algebraic extension of the prototype,} \asecfog\rev{, is also presented. Such an extension features the possibility of using different (semiring-based) trust models so to weight (and assess) the security level of eligible application deployments.}

\medskip
\noindent
The rest of this paper is organised as follows. After briefly introducing the \problog\ language (Section \ref{sec_problog}), needed to understand the proposed solution, we detail (Section \ref{sec_methodology}) the \secfog\ methodology and its implementation, while showing how it can be used on simple examples, \rev{relying on a simple default probabilistic trust model}. Afterwards, we describe a larger lifelike example of secure application deployment and illustrate how the \secfog\ prototype can be used along with other existing tools for application placement in Cloud-Edge scenarios (Section \ref{sec_example}). \rev{Then, in Section} \ref{sec_asecfog}\rev{, we discuss how} \secfog \rev{ can embed more complex trust models based on the default one, and we present an algebraic extension of the basic prototype,} \asecfog, \rev{that permits relying on arbitrary semiring-based trust models.} Finally, after discussing related work (Section \ref{related}), we draw some conclusions and point to some directions for future work (Section \ref{conclusions}).

%% file: src/_Background.tex
\section{Background: The {\sf ProbLog} Language}
\label{sec_problog}

\noindent
Being \secfog\ a declarative methodology based on probabilistic reasoning about declared infrastructure capabilities and security requirements, it was natural to prototype it by relying on probabilistic logic programming. To implement both the model and the matching strategy we used a language called \problog\ \cite{problog15}. \problog\ is a Python package that permits writing logic programs that encode complex interactions between sets of heterogeneous components, capturing the inherent uncertainties that are present in real-life situations. 

\medskip
\noindent \problog\ programs are logic programs in which some of the facts are annotated with (their) probabilities. 
\problog\ facts, such as 
\begin{Verbatim}[fontsize=\footnotesize]
    p::f.
\end{Verbatim}
\noindent represent a statement {\small\tt f} which is true with probability\footnote{A fact declared simply as {\small\tt f.} is assumed to be true with probability 1.} {\small\tt p}. \problog\ rules, like
\begin{Verbatim}[fontsize=\footnotesize]
    r :- c1, ... , cn.
\end{Verbatim}
\noindent represent a property {\small\tt r} inferred when {\small\tt c1} $\wedge\ \cdots\ \wedge$ {\small\tt cn} holds. Variable terms start with upper-case letters, constant terms with lower-case letters. Semicolon {\tt ;} can be used to express OR conditions.

\medskip\noindent
Each program defines a probability distribution over logic programs where a fact {\small\tt p::f.} is considered true with probability {\small\tt p} and false with probability $1 - {\small\tt p}$. 
The \problog\ engine \cite{problog07} determines the success probability of a query {\small\tt q} as the probability that {\small\tt q} has \textit{a} proof, given the distribution over logic programs. The engine also permits to automatically obtain a graphical representation of the AND-OR tree associated with the ground program that it used to answer a given query. This explains how the results are obtained. Finally, the \problog\ engine can be used in \textit{explanation mode} to get, for each query, the list of all mutually exclusive proofs that lead to infer it. 

%% file: src/_Methodology.tex
\section{Methodology and Implementation}
\label{sec_methodology}

\subsection{Overview}

\begin{figure}[]
\centering
  \includegraphics[width=0.99\textwidth]{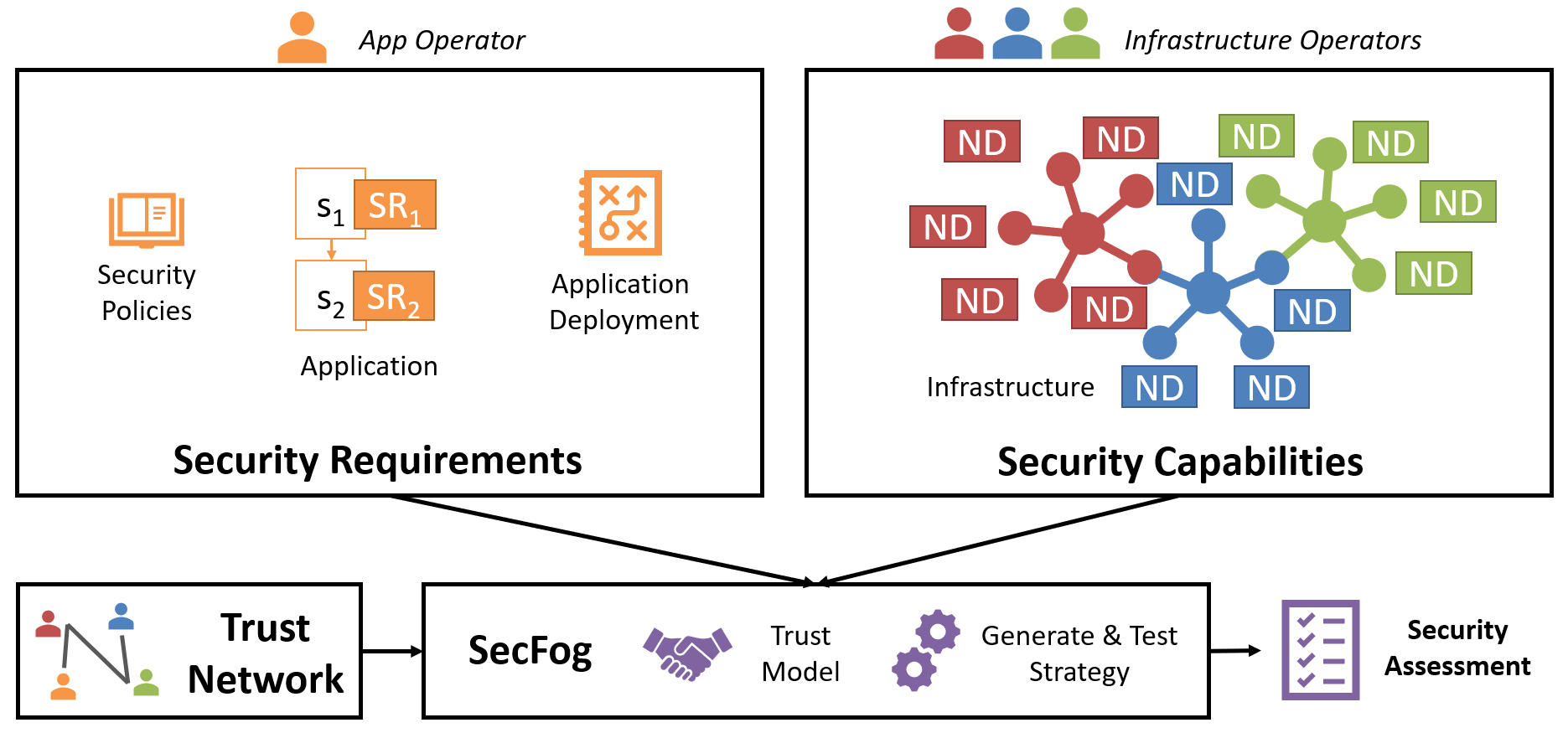}
\caption[]{Bird's-eye view of {\sf\footnotesize SecFog}.}
\label{secfog}
\end{figure}

\noindent
Figure \ref{secfog} offers an overview of the \secfog\ ingredients that will be thoroughly described in this section. First of all, \secfog\ considers two \textit{roles} for its users:

\begin{itemize}
    \item[-] \textit{infrastructure operators}, in charge of managing targeted Cloud-Edge nodes, and providing a description of the provisioned infrastructure capabilities to their users, and
    \item[-] \textit{application operators}, in charge of designing and managing application deployments by specifying their requirements (e.g., hardware, software, QoS), by monitoring deployment performance, and by re-distributing or re-configuring application services when needed.
\end{itemize}

\noindent Cloud-Edge stakeholders\footnote{Naturally, one stakeholder can play more than one role at a time and there can be more stakeholders playing the same role \cite{arcangeli}. For instance, an infrastructure operator can provide IoT, Edge, or Cloud infrastructures to its customers, whilst managing applications and services over the very same infrastructure, or an application operator can act as an infrastructure provider when sharing her home router as an Edge capability for application deployment.} might want to look for secure deployments of their applications and require explanations about such security assessment.  

As per the \secfog\ input, infrastructure operators must provide a \textit{Node Descriptor} (\ND) for each of their managed nodes, declaring all the available security capabilities
and their estimated effectiveness against attacks. These constitute the description of all \textit{Security Capabilities} in the Cloud-Edge system.

On the other hand, application operators must provide a description of their \textit{Application} and specify the \textit{Security Requirements} of each application service (\CR), in terms of the common vocabulary or, possibly, by means of a custom set of \textit{Security Policies} declared over that vocabulary. The application operator can also possibly specify a complete or partial application deployment in case she wants to assess the security level of deployments under such constraints.

Finally, each stakeholder -- infrastructure or application operator -- can declare a trust degree towards any other stakeholder, as a set of opinions. Such weighted relations contribute to a \textit{Trust Network} that is input to \secfog.

\medskip\noindent
The \secfog\ prototype, given all Security Requirements, Security Capabilities and a Trust Network, features:

\begin{enumerate}
    \item a \textit{Generate \& Test Strategy} to determine secure deployments by matching application requirements to infrastructure security capabilities (also considering their effectiveness against attacks), and
    \item a \textit{Trust Model} that is capable of completing the Trust Network, exploiting transitivity of trust chains and relying on the opinions declared by all stakeholders.
\end{enumerate}

\noindent The output of \secfog\ is a \textit{Security Assessment} of all eligible deployments determined through the exploration of the search space. For each candidate deployment, the prototype obtains its security level by multiplying the effectiveness of all exploited security capabilities, suitably weighted
by trust degrees towards the operator providing them. \rev{Trust weights are crucial to mitigate and contrast the potential presence of operators that might declare unprecise, outdated or deliberatedly false data about the security capabilities they provide.}\rev{ As better detailed in Sections} \ref{trust_implementation} and \ref{sec_asecfog}\rev{, our methodology allows to embed different trust models and to consider alternative interpretations for the values related to the effectiveness of available security capabilities. For instance, values declared by a certain provider can be overwritten by the results of \textit{objective} measurements performed directly by application operators and based on their interactions with the provider, on asset locations, on the level of hardening of different host servers.} Last but not least, each output comes with a proof that explains in a graphical, human-readable format how such result was obtained.

In the next paragraphs, we detail each of the ingredients we mentioned and, for each, we give the implementation\footnote{The {\sf\scriptsize ProbLog} code of {\sf\scriptsize SecFog} prototype and of the examples of Section \ref{sec_methodology} and \ref{sec_example} is publicly available at \url{https://github.com/di-unipi-socc/SecFog}.} as well as concrete, executable examples within our \problog\ prototype.

\subsection{Security in Cloud-Edge Computing}
\label{taxonomy_sec}

\begin{figure}
\centering
\scalebox{0.75}{
\begin{tikzpicture}[
  level 1/.style={sibling distance=35mm},
  edge from parent/.style={-,draw},
  >=latex]

\node[root] {\textbf{CLOUD-EDGE SECURITY}}
  child {node[level 2] (c1) {\textbf{Virtualisation}}}
  child {node[level 2] (c2) {\textbf{Communications}}}
  child {node[level 2] (c3) {\textbf{Data}}}
  child {node[level 2] (c4) {\textbf{Physical}}}
  child {node[level 2] (c5) {\textbf{Other}}};

\begin{scope}[every node/.style={level 3}]
\node [below of = c1, xshift=15pt] (c11) {Access Logs};
\node [below of = c11] (c12) {Authentication};
\node [below of = c12] (c13) {Host IDS};
\node [below of = c13] (c14) {Process Isolation};
\node [below of = c14] (c15) {Permission Model};
\node [below of = c15] (c16) {Resource Usage Monitoring};
\node [below of = c16] (c17) {Restore Points};
\node [below of = c17] (c18) {User Data Isolation};

\node [below of = c2, xshift=15pt] (c21) {Certificates};
\node [below of = c21] (c22) {Firewall};
\node [below of = c22] (c23) {IoT Data Encryption};
\node [below of = c23] (c24) {Node Isolation Mechanims};
\node [below of = c24] (c25) {Network IDS};
\node [below of = c25] (c26) {Public Key Cryptography};
\node [below of = c26] (c27) {Wireless Security};

\node [below of = c3, xshift=15pt] (c31) {Backup};
\node [below of = c31] (c32) {Encrypted Storage};
\node [below of = c32] (c33) {Obfuscated Storage};

\node [below of = c4, xshift=15pt] (c41) {Access Control};
\node [below of = c41] (c42) {Anti-tampering Capabilities};

\node [below of = c5, xshift=15pt] (c51) {Audit};

\end{scope}

\foreach \value in {1,2,3,4,5,6,7,8}
  \draw[-] (c1.195) |- (c1\value.west);

\foreach \value in {1,...,7}
  \draw[-] (c2.195) |- (c2\value.west);

\foreach \value in {1,...,3}
  \draw[-] (c3.195) |- (c3\value.west);
  
\foreach \value in {1,...,2}
  \draw[-] (c4.195) |- (c4\value.west);
  
  \foreach \value in {1}
  \draw[-] (c5.195) |- (c5\value.west);
\end{tikzpicture}
}
\caption[]{An example of taxonomy of security capabilities in Cloud-Edge.}
\label{taxonomy}
\end{figure}
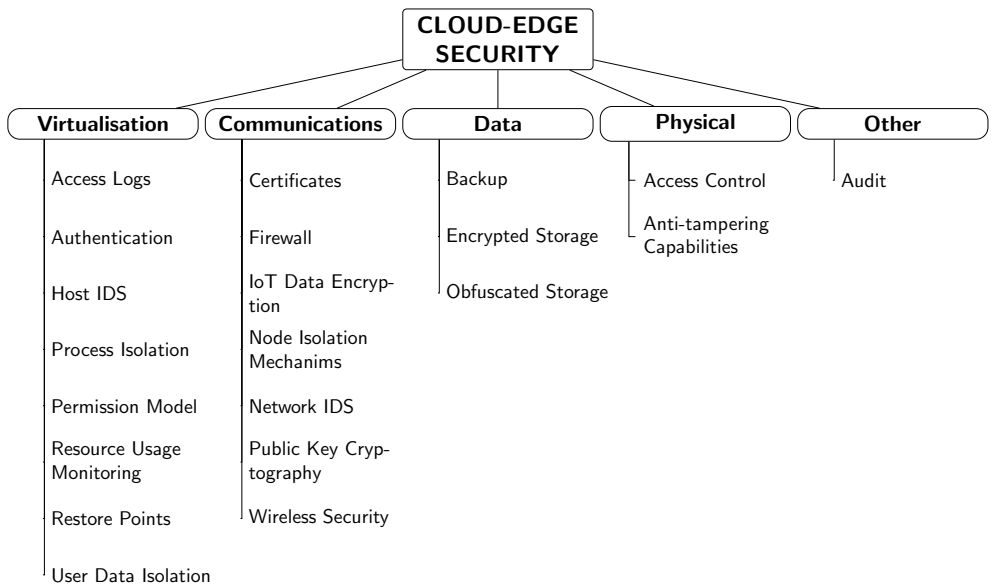

\noindent In Cloud-Edge computing infrastructures, end-to-end security must cover everything between the Cloud and the IoT. Naturally, the security requirements of multi-service applications will highly vary depending on business cases, target markets and vertical use cases as well as on the functionality offered by different components. For instance, a database storing video footage from domestic CCTV will probably require more security countermeasures to be active with respect to a database collecting outdoor temperature data from a city or neighbourhood.
The OpenFog Consortium \cite{openfog} highlighted the need for new Cloud-Edge computing platforms to guarantee privacy, anonymity, integrity, trust, attestation, verification and measurement. 
Whilst security control frameworks exist for Cloud computing scenarios (e.g., the EU Cloud SLA Standardisation Guidelines \cite{cloudslaguidelines} or the ISO/IEC 19086), to the best of our knowledge, no standard exists yet that defines security objectives for Cloud-Edge application deployments. 
Based on recent surveys about security aspects in the novel Cloud-Edge landscapes (i.e., \cite{mezni2018security}, \cite{mukherjee2017security}, \cite{ni2017securing}), we devised a simple example of taxonomy
(Figure \ref{taxonomy}) of security capabilities that can be offered by Cloud and Edge nodes and therefore used for reasoning on the security levels of given IoT application deployments. 

Security capabilities that are common with the Cloud might assume renewed importance in Cloud-Edge scenarios, due to the limited resources of the devices installed closer to the edge of the Internet. For instance, guaranteeing physical integrity of and user data isolation at an access point with Edge capabilities might be very difficult. Dually, the possibility to encrypt or obfuscate data at Edge nodes, along with encrypted IoT communication and physical anti-tampering machinery, will be key to protect those application deployments that need data privacy assurance.

In the following description of the \secfog\ methodology, we will assume that all involved parties (viz., application and infrastructure operators) share the vocabulary of the example taxonomy\footnote{Factually, different operators can employ different vocabularies and then exploit mediation \cite{mediation} mechanisms, capable of translating one into another.} in Figure \ref{taxonomy}. \rev{Naturally, in an operational system based on } \secfog\ \rev{, it will be necessary to adopt an extended and refined version of such taxonomy -- with full definitions of the considered security capabilities -- which we expect will be available as soon as normative security frameworks will get established for (Cloud-)Edge computing infrastructures.}

\subsubsection{Security Capabilities: the Infrastructure}
\label{infrastructure_implementation}

\noindent
The \textit{Infrastructure} can be simply described by infrastructure operators as a set of facts declaring a node and its security capabilities, weighted by the probability that each capability can resist against malicious attacks. Such probability represents a measure of the effectiveness of the enforced security countermeasure. Figure \ref{problog_taxonomy} lists the vocabulary of \problog\ facts that can be used to describe node capabilities in terms of the taxonomy of Figure \ref{taxonomy}.

\begin{figure}[ht]
    \Centering
        \begin{multicols}{2}
            \begin{Verbatim}[fontsize=\footnotesize, xleftmargin=-1.8cm]
                % virtualisation
                access_logs(N).
                authentication(N).
                host_ids(N).
                process_isolation(N).
                permission_model(N).
                resource_monitoring(N).
                restore_points(N).
                user_data_isolation(N).
                
                % communication
                certificates(N).
                iot_data_encryption(N).
                firewall(N).
                node_isolation_mechanism(N).
                network_ids(N).
                public_key_cryptography(N).
                wireless_security(N).
                \end{Verbatim}
            \columnbreak    
            \begin{Verbatim}[fontsize=\footnotesize, xleftmargin=-1.5cm]
                % data
                backup(N).
                encrypted_storage(N).
                obfuscated_storage(N).
                
                % physical
                access_control(N).
                anti_tampering(N).
                
                % audit
                audit(N).
            \end{Verbatim}
        \end{multicols}
    \caption{Example Cloud-Edge security capabilities in \problog.}
    \label{problog_taxonomy}
\end{figure}

\noindent 
\rev{It is worth noting that undeclared capabilities are assumed to be unavailable at the considered node. Dually, some providers might decide not to disclose the effectiveness of security capabilities against malicious attacks. In such a case, they would be unfairly considered as the best ones by the system as plainly declared facts are assumed to be always true. To prevent this from happening, application operators can for instance rely on data collected from previous interactions with the provider and specify probabilities on their own, or leverage the adoption of a convenient trust model (see Sections} \ref{trust_implementation}
 \rev{and} \ref{sec_asecfog})\rev{ to mitigate the effects of such omissions. Indeed, we reasonably expect that providers declaring unreliable data will get lower trust from application operators within the system.}

\medskip
\example\ A cloud node identified as {\small\tt cloud1} and managed by a certain infrastructure operator {\small\tt cloudOp1} can be specified in \secfog\ as:
\begin{Verbatim}[fontsize=\footnotesize]
    node(cloud1, cloudOp1).
\end{Verbatim}
\noindent In case node {\tt\small cloud1} features some form of firewall that is guaranteed to resist an attack (e.g., network traffic flood, packet fragmentation) with a likelihood of $99.99\%$, the previous line can be simply followed by:
\begin{Verbatim}[fontsize=\footnotesize]
    0.9999::firewall(cloud1).
\end{Verbatim}
Similarly, an edge node {\small\tt edge3} managed by an operator {\small\tt appOp42} and featuring a broken wireless security system like WEP and an encrypted storage considered $99\%$ effective, is declared as:
\begin{Verbatim}[fontsize=\footnotesize]
    node(edge3, appOp42).
    0.01::wireless_security(edge3).
    0.99::encrypted_storage(edge3).
\end{Verbatim}
Overall, this type of fact declarations can be used to specify the Node Descriptors by infrastructure operators, as sketched in Figure \ref{secfog}.{\footnotesize\qed}

\subsubsection{Security Requirements: the Application}
\label{application_implementation}

\noindent As aforementioned, \secfog\ enables application operators to specify an application along with the services that compose it. Based on the same common vocabulary of Figure \ref{taxonomy_sec}, application operators can then define (non-trivial) custom Security Policies that can be used to declare the Security Requirements of each service composing the application. Custom security policies can be either existing ones, inferred from the presence of certain node capabilities, or they can be autonomously specified/enriched by the application operators, depending on business-related considerations. 

\medskip
\example\ First, an application for {\tt\small smartfarming} consisting of three services {\tt\small s1}, {\tt\small s2} and {\tt\small s3} can be easily specified as:
\begin{Verbatim}[fontsize=\footnotesize]
    app(smartfarming, [s1, s2, s3]).
\end{Verbatim}
Then, the application operator can decide that a node offering {\tt\small backup} capabilities together with {\tt\small encrypted\_storage} or {\tt\small obfuscated\_storage} can be considered a {\tt\small secureStorage} provider. The custom security policy described above can be specified as:
\begin{Verbatim}[fontsize=\footnotesize]
    secureStorage(N) :- 
        backup(N), 
        (encrypted_storage(N); obfuscated_storage(N)).
\end{Verbatim}
A different stakeholder might also require the availability of a {\tt\small certificate} at the node featuring secure storage and re-define the policy as: 
\begin{Verbatim}[fontsize=\footnotesize]
    secureStorage(N) :- 
        backup(N), 
        certificate(N),
        (encrypted_storage(N); obfuscated_storage(N)).
\end{Verbatim}
The Security Requirements for service {\tt\small s2} of {\tt\small smartfarming}, which needs both {\tt\small secureStorage} (as previously declared) and {\tt\small resource\_monitoring} capabilities can then be defined as:
\begin{Verbatim}[fontsize=\footnotesize]
    securityRequirements(s2, N) :-
        secureStorage(N),
        resource_monitoring(N).
\end{Verbatim}
Overall, this permits application operators to quickly and flexibly specify all the Security Requirements of their software systems by exploiting combinations of custom Security Policies and basic security capabilities.{\footnotesize\qed}

\subsection{Generate \& Test Strategy}
\label{genandtestsection}

\noindent
The \problog\ listing in Figure \ref{genandtest} defines the declarative Generate \& Test strategy of \secfog. This strategy exploits a knowledge base of facts and rules that defines a Cloud-Edge infrastructure and its Security Capabilities along with an application and its Security Requirements, declared as we have seen in the previous section. It is worth noting that \secfog\ strategy can be used both
\begin{itemize}
    \item[(\textit{a})] to determine (or complete) secure application deployments, assessing their security level, and
    \item[(\textit{b})] to assess the security level guaranteed by complete input deployments, which may be already used by the application operators. 
\end{itemize}

\noindent
In both cases (\textit{a}) and (\textit{b}), the quantitative security assessment considers the available security capabilities and their declared effectiveness against attacks.

The rule for {\tt\small secFog(OpA, A, D)} inputs an application operator {\tt\small OpA}, an application {\tt\small A} she is in charge of deploying and a (possibly empty, or partial) deployment {\tt\small D} of such application. 
First, it checks that {\tt\small A} has been declared as an application {\tt\small app(A, L)} (line 2), then it evaluates the predicate {\tt\small deployment(OpA, L, D)} (line 3).
Recursively, for each component {\small\tt C} in the list of the application components, it checks whether it (can be) has been deployed to a node {\small\tt N} (line 7) that can satisfy the security requirements declared by the application operator for {\small\tt C}, i.e. whether {\small\tt securityRequirements(C,N)} holds (line 8).

\begin{figure}[ht]
    \Centering
            \begin{Verbatim}[fontsize=\footnotesize]
        secFog(OpA, A, D) :-                            (1)
            app(A, L),                                  (2)
            deployment(OpA, L, D).                      (3)
                                                        (4)
        deployment(_,[],[]).                            (5)
        deployment(OpA,[C|Cs],[d(C,N,OpN)|D]) :-        (6)
            node(N,OpN),                                (7)
            securityRequirements(C,N),                  (8)
            deployment(OpA,Cs,D).                       (9)
            \end{Verbatim}
    \caption{The Generate \& Test strategy of {\sf\footnotesize SecFog}.}
    \label{genandtest}
\end{figure}

\begin{figure}
    \centering
    \includegraphics[width=\textwidth]{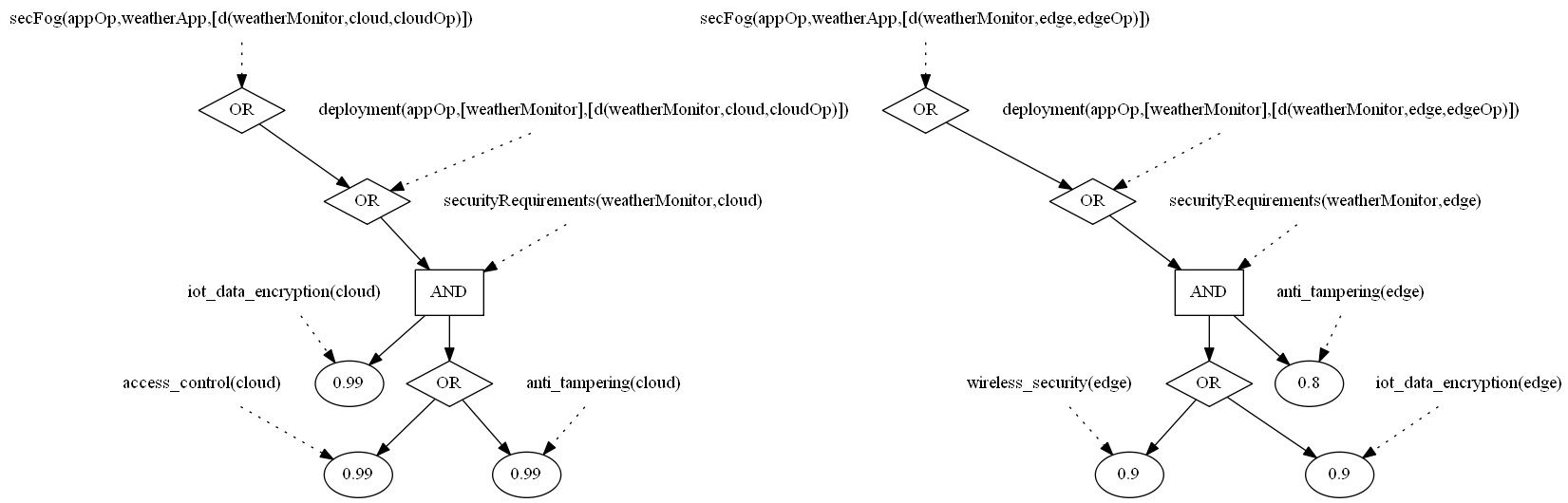}
    \caption{Graphical ground program of the {\tt\footnotesize weatherApp} example.}
    \label{weatherground}
\end{figure}

\medskip
\example\ Considering a single-service application, managing the weather data of a municipality, and an infrastructure composed of two (one Cloud and one Edge) nodes declared as follows:
\begin{Verbatim}[fontsize=\footnotesize]
    %%% Application, specified by appOp
    app(weatherApp, [weatherMonitor]).
    securityRequirements(weatherMonitor, N) :-
        (anti_tampering(N); access_control(N)),
        (wireless_security(N); iot_data_encryption(N)).
    
    %%% Cloud node, specified by cloudOp
    node(cloud, cloudOp).
    0.99::anti_tampering(cloud).
    0.99::access_control(cloud).
    0.99::iot_data_encryption(cloud).
    
    %%% Edge node, specified by edgeOp
    node(edge, edgeOp).
    0.8::anti_tampering(edge).
    0.9::wireless_security(edge).
    0.9::iot_data_encryption(edge).
\end{Verbatim}
\noindent Running the query 
\begin{Verbatim}[fontsize=\footnotesize]
    query(secFog(appOp,weatherApp,D)).
\end{Verbatim}
\noindent outputs the resulting secure deployments for the {\tt\small weatherApp}, along with a value in the range $[0,1]$ that represents their assessed security level (based on the declared effectiveness of infrastructure capabilities that are exploited by each possible deployment):
\begin{Verbatim}[fontsize=\footnotesize]
secFog(appOp,weatherApp,[d(weatherMonitor,cloud,cloudOp)]):     0.989901
  secFog(appOp,weatherApp,[d(weatherMonitor,edge,edgeOp)]):     0.792
\end{Verbatim}
The result, highlighting the deployment to the Cloud as the most secure solution, can be explained by looking at Figure \ref{weatherground}, which graphically depicts the AND-OR trees of the two ground programs that lead to the output results. Such graphical explanations can be obtained automatically, by using \problog\ in {\tt\small ground} mode\footnote{\url{https://problog.readthedocs.io/en/latest/cli.html\#grounding-ground}}. 
Note that the \problog\ engine performs an AND-OR graph search over the ground program to determine the query results.
For instance, the value associated with {\tt\small securityRequirements(weatherMonitor,cloud)} is obtained as:
{\scriptsize
\begin{multline}
p({\tt anti\_tampering(cloud)}) \times\ p({\tt iot\_data\_encryption(cloud)})\ +\\
(1-p({\tt\small anti\_tampering(cloud)})) \times\ p({\tt\small access\_control(cloud)}) \times\ p({\tt\small iot\_data\_encryption(cloud)})
= \\
.99 \times .99 +  (1-.99) \times .99 \times .99
=\\
0.989901
\end{multline}
}
\noindent As for the AND-OR graph of the ground program, also this proof can be obtained automatically, by using \problog\ in {\tt\small explain} mode\footnote{\url{https://problog.readthedocs.io/en/latest/cli.html\#explanation-mode-explain}}.
{\footnotesize\qed}

\subsection{\rev{Default} Trust Model}
\label{trust_implementation}

%
\noindent
The pervasive and highly distributed nature of new Cloud-Edge deployments imposes to deal not only with the effectiveness of the adopted security capabilities but also with the trust degrees towards various, potentially unknown, infrastructure operators. \rev{As aforementioned in Section} \ref{sec_methodology}\rev{, trust considerations are also essential both to discredit unreliable (e.g., lazy or dishonest) providers that declare unprecise, outdated or false data about their assets, and to discourage others to do so.}
In this section, we show how it is possible to \rev{smoothly} extend the prototype \secfog\ described up to now so to include a simple, yet powerful, \rev{probabilistic} trust model. \rev{In Section} \ref{sec_asecfog}\rev{, we will detail how} \secfog\ \rev{can be generalised to accommodate and leverage more complex trust models.}

The default trust model of \secfog\ considers \textit{direct trust relations} between two stakeholders $A$ and $B$ as the probability that $A$ can trust $B$, \rev{e.g.} based on an aggregate of all previous interactions they had. The proposed trust model combines such direct trust opinions from different stakeholders and completes the (possibly partial) trust network input to \secfog\ with missing indirect trust relations.
Specifically, the default trust model considers trust relations as transitive and explores network paths while aggregating the opinions (as declared by application and infrastructure operators). 
Opinions along paths are (unconditionally) combined via multiplication, opinions across paths are (monotonically) combined via addition. 
Intuitively, this causes opinions to deteriorate along paths and, when multiple opinions are available, to improve, by weighting more those paths that are more trustworthy  \cite{trust06}. 

Figure \ref{trustmodelcode} lists the \problog\ rules we used to define our \rev{default} model which is a (probabilistic) transitive closure of the trust network input to \secfog. Trivially, we assume that each stakeholder fully trust herself (line 1). Then, stakeholder {\tt\small A} can trust {\tt\small B} either directly (lines 3--4), or through a third party {\tt\small C} that directly or indirectly trusts {\tt\small B} (lines 5--7).

\begin{figure}[ht]
    \Centering
            \begin{Verbatim}[fontsize=\footnotesize]
                        trusts(X,X).        (1)
                                            (2)
                        trusts2(A,B) :-     (3)
                            trusts(A,B).    (4)
                        trusts2(A,B) :-     (5)
                            trusts(A,C),    (6)
                            trusts2(C,B).   (7)
            \end{Verbatim}
    \caption{Default trust model of {\sf\footnotesize SecFog}.}
    \label{trustmodelcode}
\end{figure}

\noindent Figure \ref{deployment_new_rule} shows how to include the trust model in the Generate and Test strategy of Figure \ref{genandtest}, namely by simply adding the condition {\tt\small trusts2(OpA, OpN)}.

\begin{figure}[ht]
    \Centering
            \begin{Verbatim}[fontsize=\footnotesize]
        deployment(_,[],[]).                            (1)
        deployment(OpA,[C|Cs],[d(C,N,OpN)|D]) :-        (2)
            node(N,OpN),                                (3)
            securityRequirements(C,N),                  (4)
            trusts2(OpA, OpN),                          (5)
            deployment(OpA,Cs,D).                       (6)
            \end{Verbatim}
    \caption{The {\tt\footnotesize deployment/3} predicate with trust.}
    \label{deployment_new_rule}
\end{figure}

\noindent We first show an example of this trust model alone, then we apply it to the example of Section \ref{genandtestsection} to illustrate its usage within \secfog.

\begin{figure}[h]
\centering
  \includegraphics[width=0.4\textwidth]{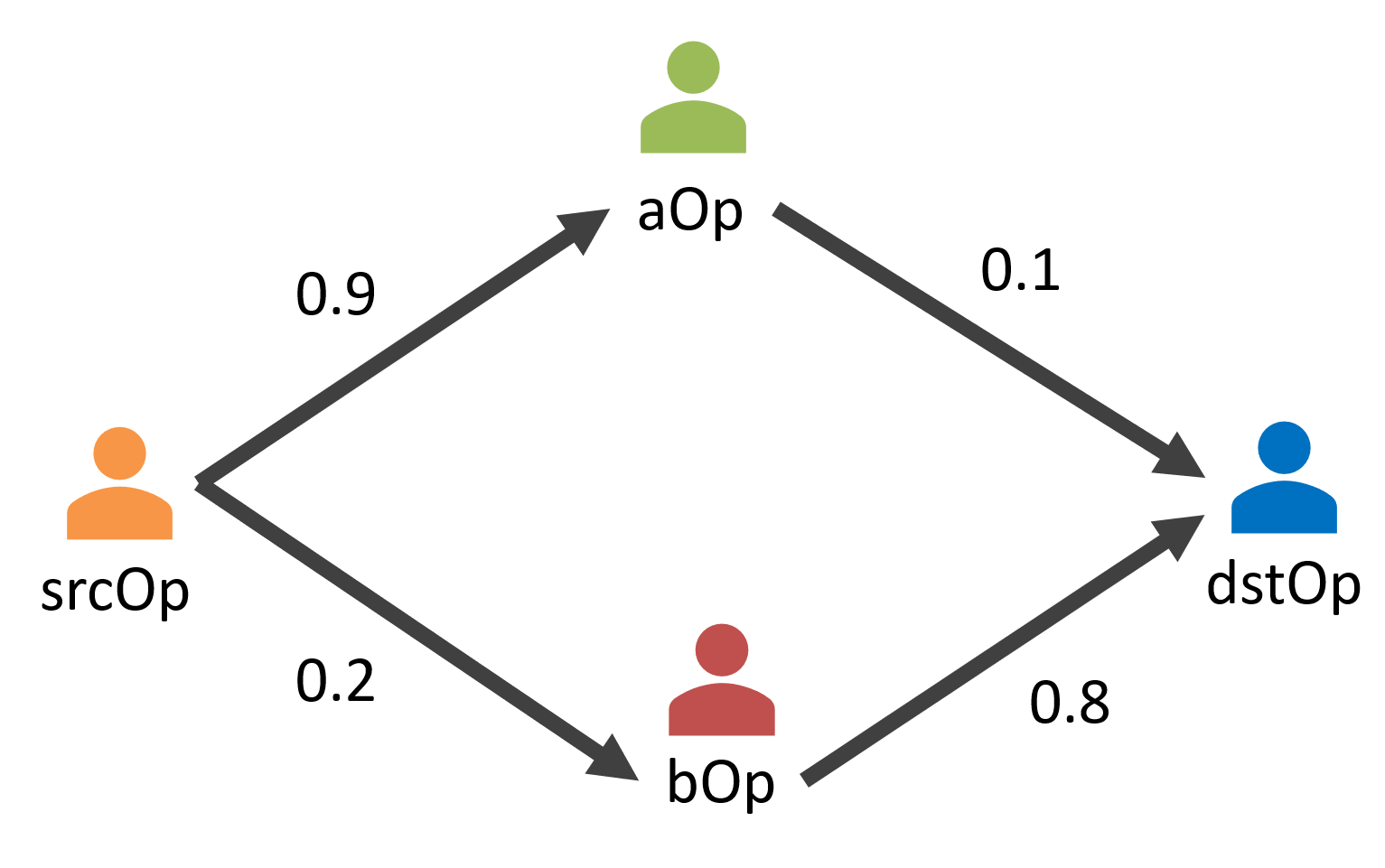}
\caption[]{A trust network.}
\label{trustnetworkexample}
\end{figure}

\medskip
\example\ Consider the trust network of Figure \ref{trustnetworkexample} and suppose to be interested in the (indirect) trust relationship between {\tt\small srcOp} and {\tt\small dstOp}.

\begin{figure}[h]
\centering
  \includegraphics[width=\textwidth]{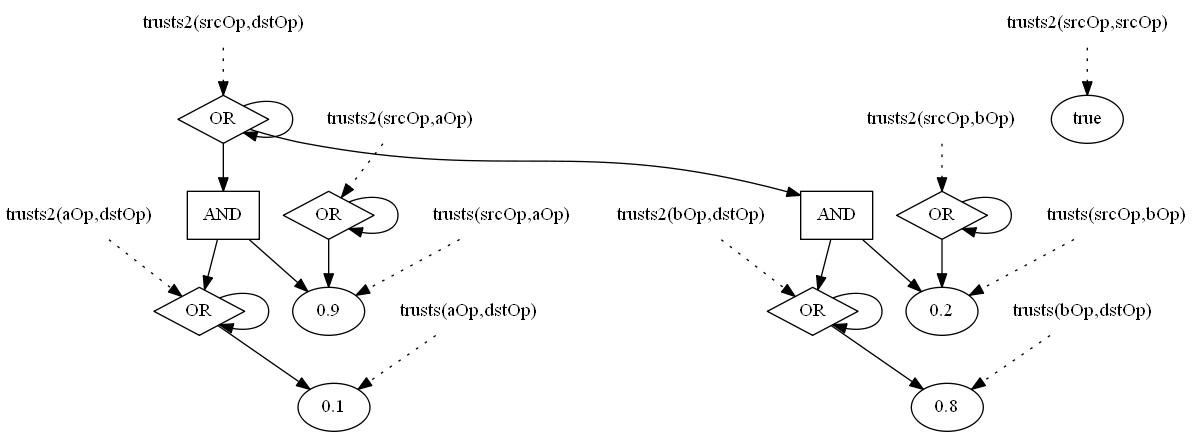}
\caption[]{Graphical ground program of the {\tt\footnotesize trusts2} example.}
\label{trustground}
\end{figure}

\noindent It can be simply computed with the \secfog\ trust model in \problog\ as:
\begin{Verbatim}[fontsize=\footnotesize]
    %%% trust relations declared by srcOp
    0.9::trusts(srcOp, aOp).
    0.2::trusts(srcOp, bOp).
    
    %%% trust relations declared by aOp
    0.1::trusts(aOp, dstOp).
    
    %%% trust relations declared by bOp
    0.8::trusts(bOp, dstOp).

    query(trusts2(srcOp, dstOp)).
\end{Verbatim}
which returns
\begin{Verbatim}[fontsize=\footnotesize]
    trusts2(srcOp,dstOp):   0.2356
\end{Verbatim}
as a result. Also in the case of the trust model, it is possible to get the ground program which explains how the final result was computed (Figure \ref{trustground}). 

\noindent From this simple example it is clear that, in the default trust model, the contribution of trust relations deteriorates along paths and that all possible paths give their contribution to the output result. Indeed, the final result corresponds to the likelihood that it is possible to establish a \textit{trust path} from {\tt\small srcOp} to {\tt\small dstOp} over the considered trust network. {\footnotesize\qed}

\begin{figure}[]
\centering
  \includegraphics[width=0.6\textwidth]{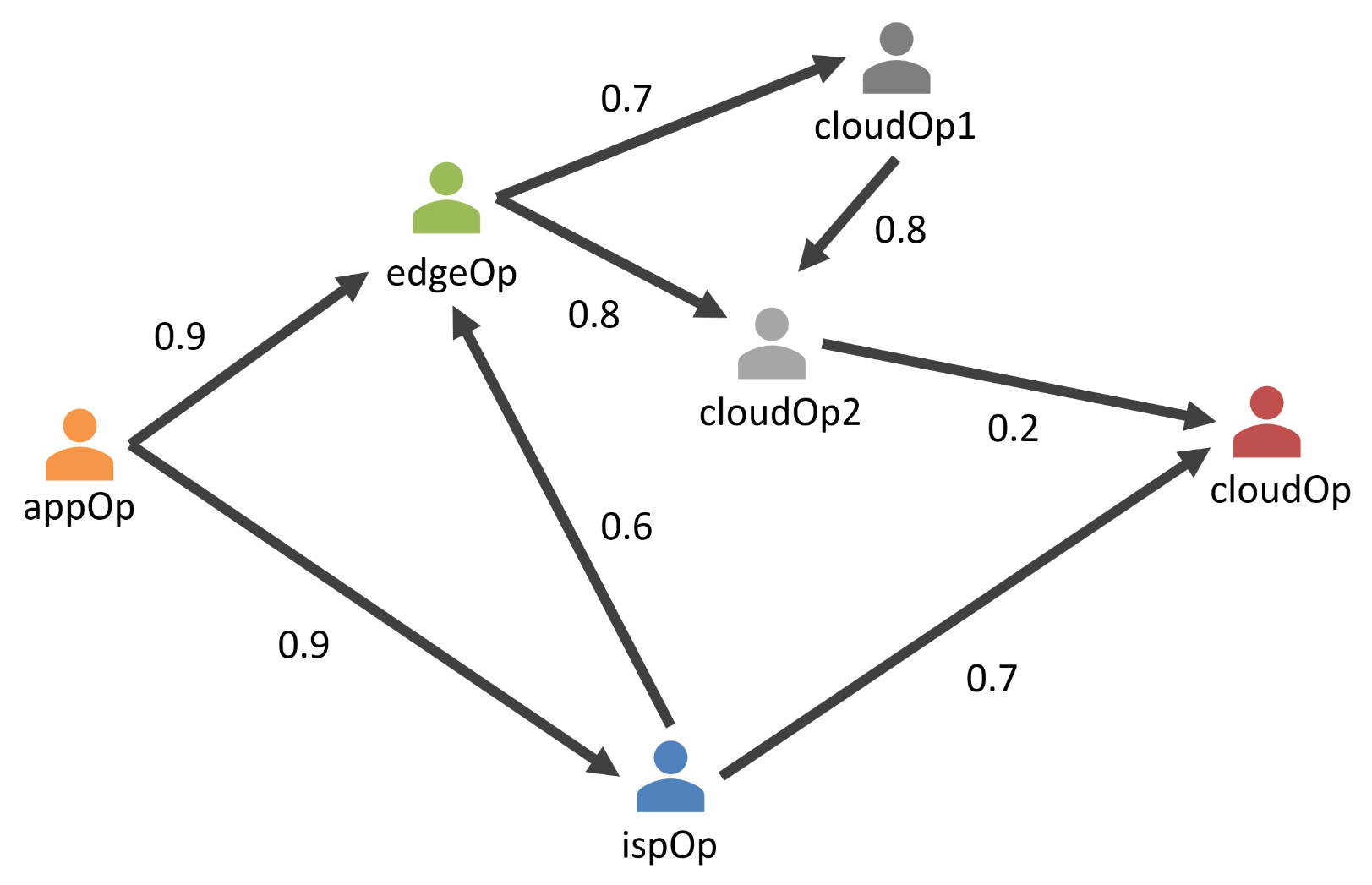}
\caption[]{Example trust network among Cloud-Edge operators.}
\label{trustnetworkexample2}
\end{figure}

\medskip
\medskip
\example\ We now retake the example of Section \ref{genandtestsection} and we solve it again by also taking into account the trust network of Figure \ref{trustnetworkexample2}.

\noindent The network is defined by the direct trust relations, which are declared by the different operators as:
\begin{Verbatim}[fontsize=\footnotesize]
    %%% trust relations declared by appOp
    .9::trusts(appOp, edgeOp).  
    .9::trusts(appOp, ispOp).
    
    %%% trust relations declared by edgeOp
    .7::trusts(edgeOp, cloudOp1).
    .8::trusts(edgeOp, cloudOp2).
    
    %%% trust relation declared by cloudOp1
    .8::trusts(cloudOp1, cloudOp2).
    
    %%% trust relation declared by cloudOp2
    .2::trusts(cloudOp2, cloudOp).
    
    %%% trust relations declared by ispOp  
    .8::trusts(ispOp, cloudOp).
    .6::trusts(ispOp, edgeOp).
\end{Verbatim}
The same query of the previous example now leads to a different result:
\begin{Verbatim}[fontsize=\footnotesize]
    secFog(appOp,weatherApp,[d(weatherMonitor,cloud,cloudOp)]):     0.76017935
      secFog(appOp,weatherApp,[d(weatherMonitor,edge,edgeOp)]):     0.755568
\end{Verbatim}
When accounting for trust, output security levels are lower than those obtained by considering only the effectiveness of security capabilities (($0.76$ and $0.75$ vs. $0.98$ and $0.79$), and the Cloud deployment does not outperform the Edge deployment anymore.
In situations like this one, the application operator might make her choice also considering other estimated non-functional parameters (e.g., cost, response time, resource usage). {\footnotesize\qed}


\medskip
\rev{Before concluding this section, it is worth noting that probabilities related to the effectiveness of security countermeasures as well as those defining trust relations can also be derived from more complex models and input to} \secfog \rev{. Indeed, the effectiveness of security countermeasures can either be extracted from the SLAs of each infrastructure provider or, alternatively, be obtained from \textit{objective} measurements and data collected by application operators from previous interactions with infrastructure providers. On the other hand, trust degrees -- which represent a \textit{subjective} piece of information -- can be either defined by the application operators or derived from more complex trust models.}

\rev{On this line,} \secfog\ \rev{can naturally embed a simplified formulation of the trust evaluation method by Tang et al.} \cite{TDLC17}. \rev{Namely,} \problog\ \rev{facts representing the security capabilities featured by a node, like}
\begin{Verbatim}[fontsize=\footnotesize]
    0.9999::firewall(cloud1). 
\end{Verbatim}
\noindent
\rev{can be exploited to represent objective trust assessments of services (computed in} \cite{TDLC17} \rev{as average conformance values between monitored and claimed QoS of a service). Problog facts representing the trust relation between stakeholders, like}
\begin{Verbatim}[fontsize=\footnotesize]
    0.9::trusts(appOp, edgeOp).
\end{Verbatim}
\noindent
\rev{can be exploited to represent subjective trust assessments of service operators (actually computed in} \cite{TDLC17} \rev{as measures of the trust of a user on a \textit{service}). A simple combination of objective and subjective trust assessment will be then directly performed by} \problog\ \rev{while evaluating the \texttt{\small deployment/3} predicate}\footnote{\rev{Please note that the purpose of the previous discussion is merely to illustrate how a simplified formulation of (the results of) the trust evaluation method of} \cite{TDLC17} \rev{can be embedded into our methodology. In fact, the trust evaluation method of} \cite{TDLC17} \rev{is more sophisticated than what we described, as it computes subjective trust assessments by taking into account both the feedback ratings of an user and those from similar, trustworthy users, and as it combines objective and subjective trust assessments by considering also their confidence.}}.

\medskip\noindent
\rev{Other simple extensions to the described probabilistic trust model consist of conditioning trust transitivity to the absence of direct trust relations, or of allowing trust transitivity in the trust network only within a specified distance (i.e. radius) from the application operator (as epitomised in Section} \ref{sec_asecfog}.

\medskip
\noindent In the next section, we exploit \secfog\ to analyse a lifelike example of IoT application deployment to Cloud-Edge infrastructure. We also make use of our \FogTorch\ prototype \cite{ccisbrogiforti} to select deployments that also meet hardware and software requirements of the example application. 

%% file: src/_Example.tex
\section{Motivating Example}
\label{sec_example}

\subsection{Infrastructure}

\begin{figure}[!ht]
\centering
\includegraphics[width=\textwidth]{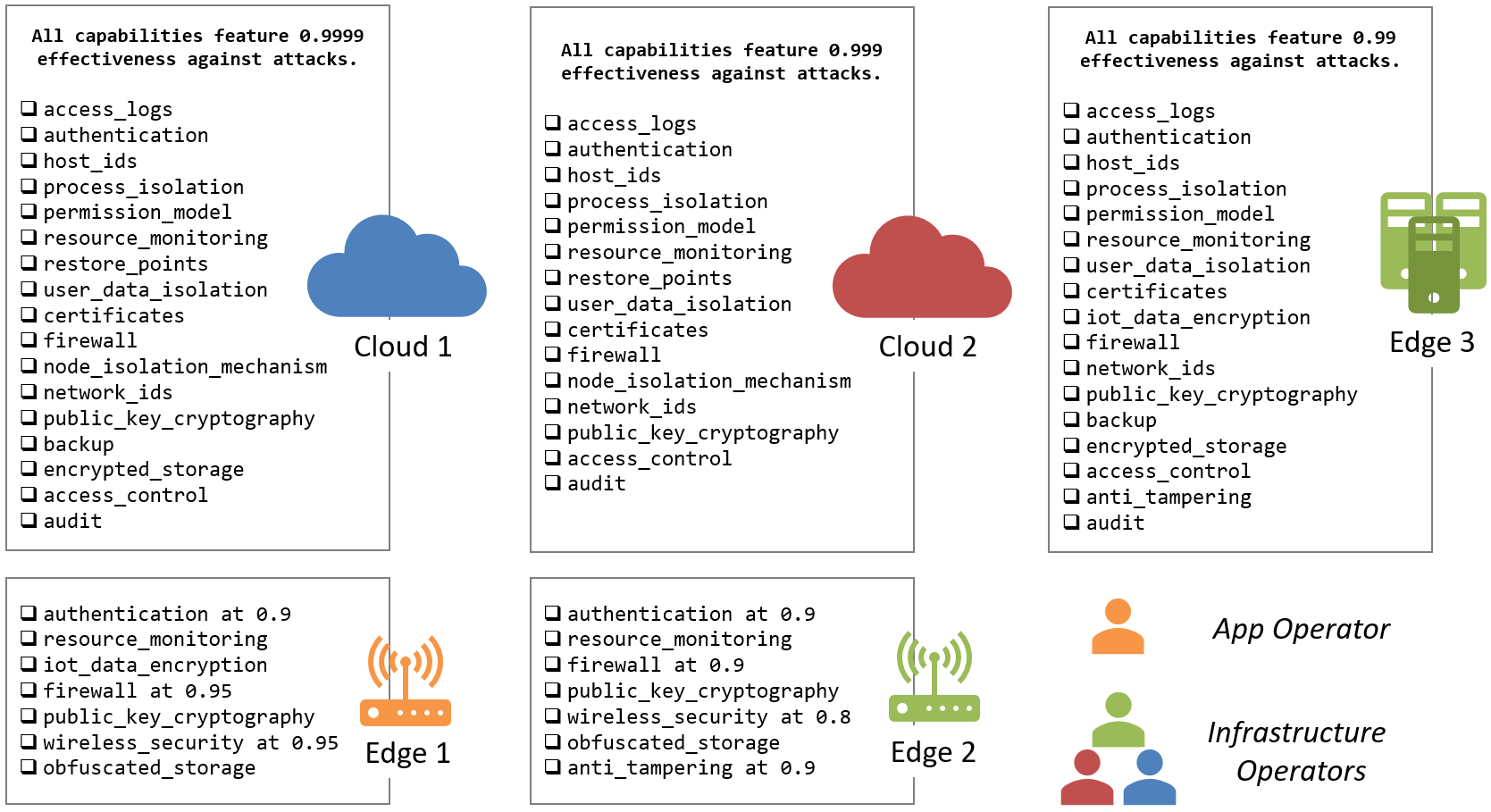}
\caption[]{Cloud-Edge infrastructure security capabilities.}
\label{infrastructure}
\end{figure}

\noindent Figure \ref{infrastructure} shows the Cloud-Edge infrastructure -- two Cloud data centres, three Edge nodes -- to which a smart building application is to be deployed. For each node, the available security capabilities and their effectiveness against attacks\footnote{In Figure \ref{infrastructure}, when the effectiveness against attacks of a capability is not indicated we assume it is considered to be $1$ by the corresponding infrastructure provider.}(as declared by the infrastructure operator) are listed in terms of the taxonomy of Figure \ref{problog_taxonomy}.

\noindent Relying on such information, node descriptors can be easily expressed by each infrastructure operator through listing ground facts, as discussed in Section \ref{infrastructure_implementation}. For instance, {\tt\small edge1} directly operated by the application operator {\tt\small appOp} is described as
\begin{Verbatim}[fontsize=\footnotesize]
    node(edge1,appOp).
    0.9::authentication(edge1).
    resource_monitoring(edge1).
    iot_data_encryption(edge1).
    0.95::firewall(edge1).
    public_key_cryptography(edge1).
    0.95::wireless_security(edge1).
    obfuscated_storage(edge1).
\end{Verbatim}
\noindent All the Node Descriptors assembled following this template form the description of the \textit{security capabilities} available in the infrastructure.

\subsection{Application}

\begin{figure}[!ht]
\centering
\includegraphics[width=0.5\textwidth]{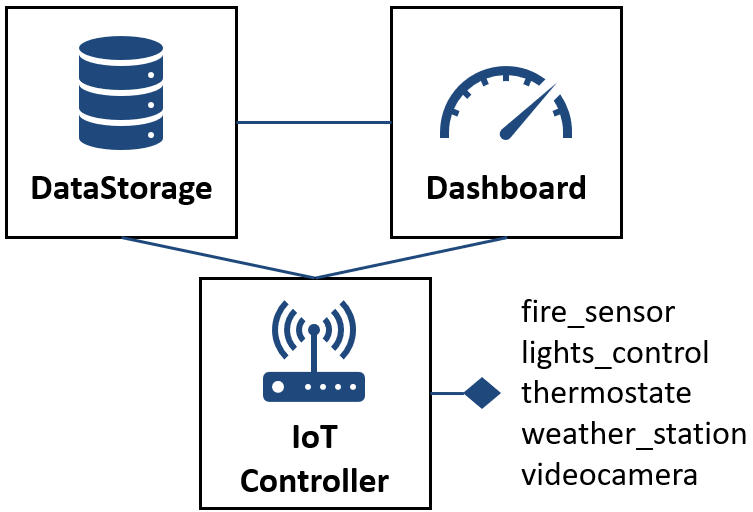}
\caption[]{Multi-service IoT application.}
\label{application}
\end{figure}

\noindent We retake the application example of \cite{ccisbrogiforti} and we extend it with security requirements. Consider a simple multi-service IoT application (Figure \ref{application}) that manages fire alarm, heating and A/C systems, interior lighting, and security cameras of a smart building. The application consists of three microservices:

\begin{itemize}
\item {\small \textsf{IoTController}}, interacting with the connected cyber-physical systems,
\item {\small \textsf{DataStorage}}, storing all sensed information for future use and employing machine learning techniques to update sense-act rules at the {\small \textsf{IoTController}} so to optimise heating and lighting management based on previous experience and/or on people behaviour, and
\item {\small \textsf{Dashboard}}, aggregating and visualising collected data and videos, as well as allowing users to interact with the system.
\end{itemize}
\noindent Each microservice represents an independently deployable component of the application \cite{microservices} and has its own security requirements. 

\noindent Particularly, application operators defined the following security requirements:

\begin{itemize}
    \item {\small \textsf{IoTController}} requires {\tt\small physical\_security} guarantees (i.e., {\small\tt access\_control} $\vee$ {\small\tt anti\_tampering}) so to avoid that temporarily stored data can be physically stolen from the deployment node,
    \item {\small \textsf{DataStorage}} requires {\tt\small secure\_storage} (viz., {\tt\small backup} $\wedge$ ({\tt\small obfuscated\_storage} $\vee$ {\tt\small encrypted\_storage})), the availability of {\tt\small access\_logs}, a {\tt\small network\_ids} in place to prevent distributed Denial of Service (dDoS) attacks, and
    \item {\small \textsf{Dashboard}} requires a {\small\tt host\_ids} installed at the deployment node (e.g., an antivirus software) along with a {\small\tt resource\_monitoring} to prevent interactions with malicious software and to detect anomalous component behaviour.
\end{itemize}

\noindent Furthermore, the application requires guaranteed end-to-end encryption among all services (viz., all deployment nodes should feature {\tt\small public\_key\_cryptography}) and that deployment nodes should feature an {\tt\small authentication} mechanism. 

\medskip
\noindent The described application and security policies translate one-to-one to the following \secfog\ clauses, as discussed in Section \ref{application_implementation}:
\begin{Verbatim}[fontsize=\footnotesize]
    %%% application
    app(smartbuilding, [iot_controller, data_storage, dashboard]).
    
    %%% security requirements
    securityRequirements(iot_controller, N) :-
        physical_security(N),
        public_key_cryptography(N),
        authentication(N).
    
    securityRequirements(data_storage, N) :-
        secure_storage(N),
        access_logs(N),
        network_ids(N),
        public_key_cryptography(N),
        authentication(N).
    
    securityRequirements(dashboard, N) :-
        host_ids(N),
        resource_monitoring(N),
        public_key_cryptography(N),
        authentication(N).
        
    %%% custom policies
    physical_security(N) :- 
        anti_tampering(N); access_control(N).
    
    secure_storage(N) :- 
        backup(N), 
        (encrypted_storage(N); obfuscated_storage(N)).
\end{Verbatim}

\subsection{Trust Network}

\begin{figure}[]
\centering
  \includegraphics[width=0.6\textwidth]{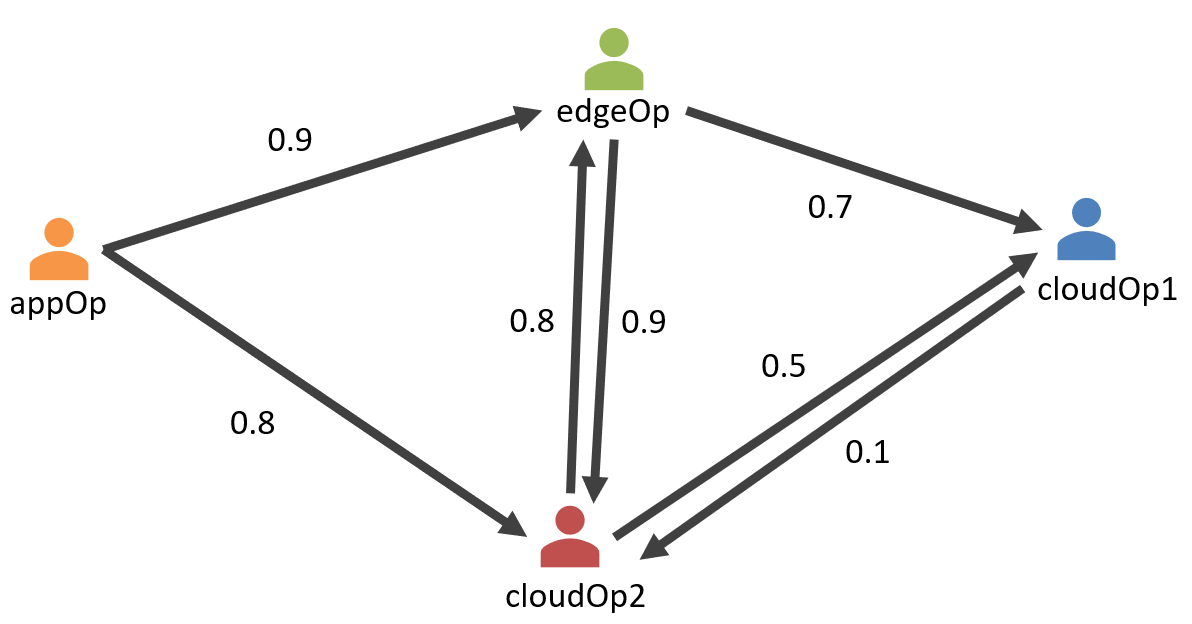}
\caption[]{Trust network of the {\footnotesize\tt smartbuilding} example.}
\label{smartbuildtrust}
\end{figure}

\noindent
Finally, as discussed in Section \ref{trust_implementation}, we consider the trust network of Figure \ref{smartbuildtrust}, which can be defined as:
\begin{Verbatim}[fontsize=\footnotesize]
    %%% trust relations declared by appOp
    0.9::trusts(appOp, edgeOp).
    0.8::trusts(appOp, cloudOp2).
    
    %%% trust relations declared by edgeOp
    0.9::trusts(edgeOp, cloudOp2).
    0.7::trusts(edgeOp, cloudOp1).
    
    %%% trust relations declared by cloudOp2
    0.1::trusts(cloudOp1, cloudOp2).
    
    %%% trust relations declared by cloudOp2
    0.8::trusts(cloudOp2, edgeOp).
    0.5::trusts(cloudOp2, cloudOp1).
\end{Verbatim}

\noindent Accounting for trust propagation, such network results in the following value of trust of {\tt\small appOp} towards infrastructure providers:
\begin{Verbatim}[fontsize=\footnotesize]
       trusts2(appOp,appOp):        1
    trusts2(appOp,cloudOp1):        0.8247
    trusts2(appOp,cloudOp2):        0.96326
      trusts2(appOp,edgeOp):        0.964
\end{Verbatim}

\subsection{Security Assessment}

\noindent As discussed in Section \ref{genandtestsection}, the \secfog\ prototype can be used to find all deployments that satisfy the security requirements of the example application to the given infrastructure, by simply issuing the query:

\vspace{-1mm}
\begin{Verbatim}[fontsize=\footnotesize]
    query(secFog(appOp, smartbuilding, D)).
\end{Verbatim}
\vspace{-1mm}

\noindent As shown in Table \ref{securityassessment}, relying on \problog\ out-of-the-box algorithms, \secfog\ prototype returns answers to the query along with a value in $[0,1]$ that represents the aggregate \textit{security level} of the inferred facts, i.e. the probability that a deployment can be considered secure according to the declared reliability of the infrastructure capabilities and to the trust degree of the application operator towards each exploited infrastructure operator.

\begin{table}[h]
\centering
{\footnotesize
\begin{tabular}{|c|c|c|c|c|c|c|}
\hline
Dep. ID  & \textbf{{\sf IoTController}}  & \textbf{{\sf DataStorage}} & \textbf{{\sf Dashboard}} & {Security} 
 \\ \hline
$\Delta$1 & {{\sf Cloud 1}} & {{\sf Cloud 1}} & {{\sf Cloud 1}} & 0.82 
\\ \hline
$\Delta$2 & {{\sf Cloud 1}} & {{\sf Cloud 1}} & {{\sf Cloud 2}} & 0.81 
\\ \hline
$\Delta$3 & {{\sf Cloud 1}} & {{\sf Cloud 1}} & {{\sf Edge 3}} & 0.78 
\\ \hline
$\Delta$4 & {{\sf Cloud 1}} & {{\sf Edge 3}} & {{\sf Cloud 1}} & 0.77
\\ \hline
$\Delta$5 & {{\sf Cloud 1}} & {{\sf Edge 3}} & {{\sf Cloud 2}} & 0.75
\\ \hline
$\Delta$6 & {{\sf Cloud 1}} & {{\sf Edge 3}} & {{\sf Edge 3}} & 0.75 
\\ \hline
$\Delta$7 & {{\sf Cloud 2}} & {{\sf Cloud 1}} & {{\sf Cloud 1}} & 0.81
\\ \hline
$\Delta$8 & {{\sf Cloud 2}} & {{\sf Cloud 1}} & {{\sf Cloud 2}} & 0.81
\\ \hline
$\Delta$9 & {{\sf Cloud 2}} & {{\sf Cloud 1}} & {{\sf Edge 3}} & 0.77
\\ \hline
$\Delta$10 & {{\sf Cloud 2}} & {{\sf Edge 3}} & {{\sf Cloud 1}} & 0.75
\\ \hline
$\Delta$11 & {{\sf Cloud 2}} & {{\sf Edge 3}} & {{\sf Cloud 2}} & 0.89
\\ \hline
$\Delta$12 & {{\sf Cloud 2}} & {{\sf Edge 3}} & {{\sf Edge 3}} & 0.87
\\ \hline
$\Delta$13 & {{\sf Edge 2}} & {{\sf Cloud 1}} & {{\sf Cloud 1}} & 0.66
\\ \hline
$\Delta$14 & {{\sf Edge 2}} & {{\sf Cloud 1}} & {{\sf Cloud 2}} & 0.65
\\ \hline
$\Delta$15 & {{\sf Edge 2}} & {{\sf Cloud 1}} & {{\sf Edge 3}} & 0.63 
\\ \hline
$\Delta$16 & {{\sf Edge 2}} & {{\sf Edge 3}} & {{\sf Cloud 1}} & 0.62
\\ \hline
$\Delta$17 & {{\sf Edge 2}} & {{\sf Edge 3}} & {{\sf Cloud 2}} & 0.72
\\ \hline
$\Delta$18 & {{\sf Edge 2}} & {{\sf Edge 3}} & {{\sf Edge 3}} & 0.72
\\ \hline
$\Delta$19 & {{\sf Edge 3}} & {{\sf Cloud 1}} & {{\sf Cloud 1}} & 0.80
\\ \hline
$\Delta$20 & {{\sf Edge 3}} & {{\sf Cloud 1}} & {{\sf Cloud 2}} & 0.78
\\ \hline
$\Delta$21 & {{\sf Edge 3}} & {{\sf Cloud 1}} & {{\sf Edge 3}} & 0.78
\\ \hline
$\Delta$22 & {{\sf Edge 3}} & {{\sf Edge 3}} & {{\sf Cloud 1}} & 0.77
\\ \hline
$\Delta$23 & {{\sf Edge 3}} & {{\sf Edge 3}} & {{\sf Cloud 2}} & 0.89
\\ \hline
$\Delta$24 & {{\sf Edge 3}} & {{\sf Edge 3}} & {{\sf Edge 3}} & 0.89
\\ \hline
\end{tabular}
}
\caption[]{Eligible deployments of the example application.}
\label{securityassessment}
\end{table}

\noindent If the application operator is only considering security as a parameter to lead her search, she would try to maximise the obtained metric and, most probably, select one among $\Delta$11, $\Delta$12, $\Delta$23, $\Delta$24. However, security might need to be considered along with other parameters so to find a suitable trade-off among them. 

In this regards, it is interesting to see how \secfog\ prototype can be used in synergy with other tools that perform multi-service application placement in Cloud-Edge scenarios. For instance, our \FogTorch\ prototype \cite{book} finds eligible deployments that guarantee software, hardware and network QoS requirements. For each deployment, it outputs the QoS-assurance (i.e., the likelihood it will meet network QoS requirements), an aggregate measure of Edge resource consumption, and an estimate of its monthly operational cost. It then employs a simple multi-objective optimisation to rank the deployments and to decide which are the better candidates.

\noindent For the {\tt\small smartbuilding} example we are analysing, among the deployments of Table \ref{securityassessment}, \FogTorch\ suggests only $\Delta$13, $\Delta$16 and $\Delta$22 \cite{ccisbrogiforti}. Naturally, with the aim of maximising security whilst considering all other requirements, the application operator would likely choose $\Delta$22.

%% file: src/_asecfog.tex
\section{Algebraic Extensions of {\sf SecFog}}
\label{sec_asecfog}

\noindent
\rev{The default probabilistic trust model of} \secfog\ \rev{shows two main limitations. Namely, it is \textit{unconditionally} \textit{ transitive} (i.e., if $A$ trusts $B$ and $B$ trusts $C$ then $A$ trusts $C$) and \textit{monotonic} (all paths towards a certain provider $P$ contribute increasing the trust degree towards it).}

\rev{More often, trust has been considered only \textit{conditionally} transferable and -- being a subjective phenomenon -- usually non-monotonically increasing} \cite{yan2007trust}. \rev{As we will show in this section, the default trust model of} \secfog\ \rev{can be easily generalised so to accommodate semiring-based trust models such as those proposed by} Theodorakopoulos and Baras \cite{trust06}, Bistarelli et al. \cite{bistarelli2008weighted,bistarelli2010semiring} or Gao et al. \cite{gao2016star}, \rev{in fields other than multi-service application deployment.} 

\rev{After briefly reviewing the mathematical definition of semirings and describing the algebraic extension of} \problog\ \rev{by Kimmig et al.} \cite{kimmig2011algebraic} \rev{we show how} \secfog\ \rev{can be generalised into an algebraic} \asecfog \rev{ and embed different semiring-based trust models. We then exploit two among those models to solve again the motivating example of Section }\ref{sec_example}.

\medskip
\noindent
\rev{A (commutative) \textit{semiring} is an algebraic structure consisting of a 5-tuple:}

$$\langle \mathcal{S}, \oplus, \otimes, {\bf 0}, {\bf 1} \rangle$$

\noindent \rev{where $\mathcal{S}$ is a set of elements, and $\oplus$ and $\otimes$ are two binary operators defined over such elements such that:}

\begin{itemize}
    \item[--] \rev{$\oplus$ is commutative and associative, and ${\bf 0}$ is its neutral element,}
    \item[--] \rev{$\otimes$ is associative, distributes over $\oplus$, and ${\bf 1}$ and ${\bf 0}$ are its neutral and absorbing elements, respectively.}
\end{itemize}

\noindent \rev{For instance, the embedded model of} \problog\ \rev{corresponds to the probability semiring}
$$\langle \mathbb{R} \cap [0,1], +, \times, 0, 1 \rangle$$

\noindent \rev{where $+$ and $\times$ denote classical addition and multiplication over real numbers.}

\rev{Intuitively, a} \problog\ \rev{program leverages input probability distributions to analyse all possible Prolog programs (i.e., worlds) that could be generated according to them. 
Assuming that $\Omega(q)$ is the set of possible worlds $W$ that entail a valid proof for a certain query $q$ 
(i.e., $\Omega(q) = \{W \mid W \models q\}$), the} \problog\ \rev{engine computes the probability $p(q)$ that $q$ holds as}

$$p(q) \ = \ \sum_{\substack{W \in \Omega(q)}} \ \prod_{f \in W} \ p(f)$$

\noindent \rev{where $f$ are facts within a certain possible world, and $p(f)$ is the probability they are labelled with. More generally, the algebraic extension of} \secfog\ \rev{allows programmers to rely on arbitrary semirings and labelling functions for computing output results.}
\rev{Intuitively, given an arbitrary semiring $\langle \mathcal{S}, \oplus, \otimes, {\bf 0}, {\bf 1} \rangle$ as defined above and a labelling function $\alpha(f)$ over the program literals, then the labelling for query $q$ is obtained as}

$$A(q) \ = \ \bigotimes_{\substack{W \in \Omega(q)}} \ \bigotimes_{f \in W} \ \alpha(f)$$

\noindent \rev{where worlds in $\Omega$ represent the set of \textit{interpretations} where $q$ is true. For all details on the algebraic extension of} \problog\ \rev{we refer the reader to} \cite{kimmig2011algebraic}.

\medskip\noindent
\rev{In what follows, we exploit such feature of} \problog\ \rev{to rely on different trust models and solve the motivating example of Section} \ref{sec_example}.

\medskip
\example\ \rev{We start by embedding in} \asecfog\ \rev{the trust model proposed in} \cite{trust06} and \cite{bistarelli2008weighted}. \rev{Still relying on the transitive closure of the trust network, their model exploits a semiring where trust is represented by couples $\langle t, c \rangle$ in the set $\mathcal{S} = (\mathbb{R}\cap [0,1]) \times (\mathbb{R}\cap [0,1])$ where $t$ represents a trust value and $c$ the confidence in such trust value assignment, i.e. the \textit{quality} of the declared opinion. Then, $\otimes$ (with neutral element $\langle 1, 1 \rangle$) is defined as}
$$\langle t, c \rangle \otimes \langle t', c' \rangle = \langle t\times t', c\times c' \rangle$$

\noindent \rev{and $\oplus$ (with neutral element $\langle 0, 0 \rangle$) is defined as}

$$
\langle t, c \rangle \oplus \langle t', c' \rangle = 
\begin{cases} 
\langle t, c \rangle & \mbox{if } c > c'\\ 
\langle t', c' \rangle & \mbox{if } c' > c \\
\langle \max \{t, t'\}, c \rangle , & \mbox{if } c = c'
\end{cases}
$$

\noindent  \rev{This model overcomes the limitations of the default trust model as it conditions trust transitivity to the confidence values and avoids monotonicity by bounding the trust value to the maximum declared one.

When embedding this trust model in} \asecfog,\rev{ we obtain that effectiveness of security countermeasures can be declared by infrastructure providers as in}
\begin{Verbatim}[fontsize=\footnotesize]
    (0.9999,1)::firewall(cloud1).
\end{Verbatim}
\noindent \rev{and that trust relations take the form}
\begin{Verbatim}[fontsize=\footnotesize]
    (0.9,0.8)::trusts(appOp, edgeOp).
\end{Verbatim}
\noindent \rev{We run} \asecfog\ \rev{with this trust model over the motivating example of Section} \ref{sec_example}, \rev{assuming that declared trust opinions are associated with confidence values as follows:}
\begin{Verbatim}[fontsize=\footnotesize]
    %%% trust relations declared by appOp
    (0.9,0.9)::trusts(appOp, edgeOp).
    (0.8,0.9)::trusts(appOp, cloudOp2).
    
    %%% trust relations declared by edgeOp
    (0.9,0.9)::trusts(edgeOp, cloudOp2).
    (0.7,0.5)::trusts(edgeOp, cloudOp1).
    
    %%% trust relations declared by cloudOp2
    (0.1,0.9)::trusts(cloudOp1, cloudOp2).
    
    %%% trust relations declared by cloudOp2
    (0.8,0.7)::trusts(cloudOp2, edgeOp).
    (0.5,0.7)::trusts(cloudOp2, cloudOp1).
\end{Verbatim}

\noindent \rev{Querying \texttt{deployment/3} results in $\Delta13$ having an overall security level of $\langle 0.116, 0.357 \rangle$, $\Delta16$ having an overall security level of $\langle 0.247, 0.510 \rangle$ and $\Delta22$ having an overall security level of  $\langle 0.296, 0.510 \rangle$.

$\Delta22$ is the most likely deployment to be chosen by the application operators as it is ranked first also with the introduction of the new model.
Akin results are obtained when substituting the $\max$ with the $\min$ operator in the semiring, which makes the computation of the security level less optimistic when confidence values coincide.}
{\footnotesize\qed}

\medskip
\example\ \rev{Building on the previous model, Gao et al.} \cite{gao2016star} \rev{recently proposed a more sophisticated one, capable of considering both trust (i.e., positive preference) and distrust (i.e., negative preference) relations. Their model also relies on a semiring where (dis)trust is represented by couples $\langle t, c \rangle$ in the set $\mathcal{T} = (\mathbb{R}\cap [-1,1]) \times (\mathbb{R}\cap [0,1])$, where $t$ represents a (dis)trust value and $c$ the confidence in such value assignment. Distrust ranges in $[-1, 0)$ and trust in $(0, 1]$, having $0$ represent an \textit{indifferent} opinion. Then, $\otimes$ (with neutral element $\langle 1, 1 \rangle$) is defined as}

$$
\langle t, c \rangle \otimes \langle t', c' \rangle = 
\begin{cases} 
\langle 0, c\times c' \rangle & \mbox{if } t < 0  \mbox{ and } t' < 0 \\
\langle t\times t', c\times c' \rangle & \mbox{otherwise} \\ 
\end{cases}
$$

\noindent \rev{and $\oplus$ (with neutral element $\langle 0, 0 \rangle$) is defined as}

$$
\langle t, c \rangle \oplus \langle t', c' \rangle = 
\begin{cases} 
\langle t, c \rangle & \mbox{if } c > c'\\ 
\langle t', c' \rangle & \mbox{if } c' > c \\
\langle \mbox{sign}(t+t')\cdot\max \{t, t'\}, c \rangle , & \mbox{if } c = c'
\end{cases}
$$

\noindent \rev{where sign$(x)$ returns $1$ if $x \geqslant 0$ and $-1$ otherwise.} 

\rev{This trust model therefore permits to express both trust and distrust opinions. It is worth mentioning that the $\otimes$ operator sets trust to $0$ when both considered opinions represent distrust along a path, what stops trust transitivity. Also, the authors of} \cite{gao2016star} \rev{propose to impose a maximum radius \texttt{D} for propagating (dis)trust relations, which can be easily obtained in} \secfog\ \rev{by revising the \texttt{trust2} predicate as listed in Fig.} \ref{fig:trustwithradius}.

\begin{figure}[ht]
    \Centering
            \begin{Verbatim}[fontsize=\footnotesize]
                    trusts2(A,B) :- trusts2(A,B,3).     (1)
                    trusts2(A,B,D) :-                   (2)
                        D > 0,                          (3)
                        trusts(A,B).                    (4)
                    trusts2(A,B,D) :-                   (5)
                        D > 0,                          (6)    
                        trusts(A,C),                    (7)
                        NewD is D - 1,                  (8)
                        trusts2(C,B,NewD).              (9)    
            \end{Verbatim}
    \caption{The {\tt\footnotesize trust2/3} predicate with maximum propagation radius \texttt{D}$=3$.}
    \label{fig:trustwithradius}
\end{figure}

\noindent \rev{We now run the motivating example with the new trust model and \texttt{D} set to $3$, assuming that Cloud providers decide to declare the following distrust opinions towards each other (instead of the trust opinions previously specified):}
\begin{Verbatim}[fontsize=\footnotesize]
    (-0.1,0.9)::trusts(cloudOp1, cloudOp2).
    (-0.1,0.7)::trusts(cloudOp2, cloudOp1).
\end{Verbatim}
\rev{As a result, the three eligible deployments $\Delta13$, $\Delta16$ and $\Delta22$ obtain security levels of $\langle 0.005, 0.357 \rangle$, $\langle 0.05, 0.510 \rangle$ and $\langle 0.06, 0.510 \rangle$ respectively. Despite $\Delta22$ ranks first, such results, very close to $0$ (i.e. to an indifferent opinion), might induce the application operators to consider upgrading part of the infrastructure they manage so to permit deployment to their assets, or to include other providers in their analysis.} {\footnotesize\qed}

%% file: src/_Related.tex
\section{Related Work}
\label{related}

\noindent
Among the studies focussing on the placement of multi-service applications to Cloud nodes, very few approaches considered security aspects when determining eligible application deployments, mainly focussing on improving performance, resource usage and deployment cost \cite{kaur2017taxonomy,mezni2018security}, or on identifying potential data integrity violations based on pre-defined risk patterns \cite{zoltan}. Other existing research considered security mainly when treating the deployment of business processes to (federated) multi-Clouds (e.g., \cite{nacer2016obfuscating,goettelmann2014security,wen2017cost}). Similar to our work, Luna et al. \cite{luna} were among the first to propose a quantitative reasoning methodology to rank single Cloud providers based on their security SLAs, and with respect to a specific set of (user-weighted) security requirements. Recently, swarm intelligence techniques \cite{mezni2018security} have been exploited to determine eligible deployments of composite Cloud applications, considering a risk assessment score based on node vulnerabilities. However, none of these works embedded trust models to consider the trust relations and the opinions of the involved stakeholders when determining secure deployments.

\medskip\noindent
\noindent Cloud-Edge computing introduces new challenges, mainly due to its pervasive geo-distribution and heterogeneity, need for QoS-awareness, dynamicity and support to interactions with the IoT, that were not thoroughly studied in previous works addressing the problem of application deployment to the Cloud \cite{icfecreview,recentreview}. 
%
Among the first proposals investigating these new lines, \cite{004} proposed a Cloud-Edge search algorithm as a first way to determine an eligible deployment of (multi-component) DAG applications to tree-like infrastructures. Their placement algorithm attempts the placement of services \textit{Edge-to-Cloud} by considering hardware capacity only. An open-source simulator -- iFogSim -- has been released to test the proposed policy against Cloud-only deployments. Building on top of iFogSim, \cite{021} triedto guarantee the application service delivery deadlines and to optimise computational resource exploitation. 
Also \cite{032} used iFogSim to implement an algorithm for optimal online placement of application components, with respect to load balancing. 
Recently, exploiting iFogSim, \cite{101} proposed a distributed search strategy to find the best service placement in Cloud-Edge infrastructures, which minimises the distance between the clients and the most requested services, based on request rates and available free resources. 
In our previous work, we also proposed a model and algorithms to determine eligible deployments of IoT applications to Fog infrastructures \cite{011} based on hardware, software and network QoS requirements. Our prototype -- \FogTorch\ --  implements those algorithms and permits to estimate the QoS-assurance, the resource consumption in the Fog layer \cite{012} and the monthly deployment cost~\cite{closer18} of the output eligible deployments.
\cite{062,027} proposed (linearithmic) heuristic algorithms that attempt deployments prioritising placement of applications to devices that feature with less free resources.

\medskip\noindent From an alternative viewpoint, \cite{001} gave a Mixed-Integer Non-Linear Programming (MINLP) formulation of the problem of placing application services aiming at satisfying end-to-end delay constraints. The problem is then solved by linearisation into a Mixed-Integer Linear Programming (MILP), showing potential improvements in latency, energy consumption and costs for routing and storage that the Cloud-Edge interplay might bring. 
Also \cite{024} adopted an ILP formulation of the problem of allocating computation to Cloud and Edge nodes so to optimise time deadlines on application execution. A simple linear model for Cloud costs is also taken into account.  
Similar solutions were proposed, attempting to optimise various metrics such as access latency, resource usage, energy consumption or data migrations cost \cite{034,017,095,061,002,019}.
\cite{107} described instead a fuzzy QoE extension of iFogSim -- based on an ILP modelling of users expectation -- which achieved improvements in network conditions and service QoS.
\noindent Regrettably, none of the discussed ILP/MILP approaches came with the code to run the experiments. 
  Conversely, \cite{030} proposed a software platform to support optimal application placement in Cloud-Edge landscapes. Envisioning resource, bandwidth and response time constraints, they compare a Cloud-only, an Edge-only or a Cloud-to-Edge deployment policy. Additionally, the authors of \cite{038} released an open-source extension of Apache Storm that performs service placement while improving the end-to-end application latency and the availability of deployed applications. 
Dynamic programming (e.g., \cite{006}), genetic algorithms (e.g., \cite{024,geneticalgorithms2019}) and deep learning (e.g., \cite{008}) were exploited promisingly in some recent works.  Overall, to the best of our knowledge, none of the previous work in the field of application placement included the possibility to look for secure deployments in Cloud-Edge scenarios, based on application requirements and infrastructure capabilities.

\medskip\noindent
When it comes to trust models and trust management \cite{trustmanagementardagna}, other works such as \cite{bistarelli2010semiring, bistarelli2008weighted} employ (weighted) logic programming and consider networks of trust (and their closures) with values in the range $[0,1]$ to express trust relations. Also \cite{li2002design} relies on logic programming to define a trust framework for role-based access policies. As in \secfog, such relations describe the belief of one stakeholder to trust another, based of the interactions they previously had.  In line with other trust models \cite{ziegler2005propagation,twigg2003attack, trust06}, we aggregate multiple trust paths.
In the context of {ad-hoc} networks, much work was done to devise certification based trust models and protocols to spread trust opinions at runtime \cite{omar2012certification}. Recently, certification-based schemes were proposed also for the Cloud scenario as in the works by Anisetti et al. \cite{anisetti2014certification, anisetti2017semi}. 

\rev{Trust and customer feedback have been employed in the definition of some cloud service selection approaches. For instance, Ding et al.} \cite{DWWO17} \rev{proposed a ranking prediction method for personalized cloud service selection, which takes into account the customer's attitude and expectation towards quality of service, and exploits collaborative filtering techinques by calculating similarities between customers. Qu et al. }\cite{QWOLLB15} \rev{proposed a context-aware and credible cloud service selection mechanism based on aggregating subjective assessments extracted from ordinary cloud consumers and objective assessments from quantitative performance testing parties. Tang et al.} \cite{TDLC17} \rev{later proposed a sophisticated trust evaluation method for cloud service selection, where \textit{objective} trust assessments (based on QoS monitoring) and \textit{subjective} trust assessments (based on  user feedback ratings) of cloud services are suitably combined.}
%


%% file: src/_Conclusions.tex
\section{Concluding Remarks}
\label{conclusions}

\noindent
In this paper, we proposed a declarative methodology, \secfog, which can be used to quantitatively assess the security level of multi-service application deployments to Cloud-Edge infrastructures. With a prototype implementation in \problog, we have shown how \secfog\ helps application operators in determining secure deployments based on specific application requirements, available infrastructure capabilities, and considering trust degrees in different Edge and Cloud providers. 

\medskip\noindent
To the best of our knowledge, \secfog\ constitutes a first well-founded, efficient and explainable effort towards such direction. The well-foundedness and efficiency of \secfog\ are guaranteed by the state-of-the-art resolution algorithms implemented within the \problog\ engine. The possibility of explaining the obtained security assessment also derives from \problog\ functionalities that allow the users to obtain graphical ground programs and proofs for the results of their queries. The \secfog\ prototype can be fruitfully used with other tools for application deployment so to identify suitable trade-offs among the estimated security level and other deployment performance indicators (e.g., QoS-assurance, resource usage, monthly cost, energy consumption), as we have shown with our prototype \FogTorch.

\medskip
\noindent
As future work, we plan to:
\begin{itemize}
    \item[-] prototype a language-based approach (as in \cite{reviewersuggestion}) and a GUI \rev{to ease the declaration of application security requirements} and to provide a user-friendly view of the recommended deployment(s), by suitably highlighting how the application security requirements are satisfied,
    \item[-] extend such GUI with a visual explanation of the reasons why a given deployment is {\it not} recommended by \secfog, and
    \item[-] engineer and integrate \secfog\ with \FogTorch\ and show their applicability to actual use cases.
\end{itemize}

\noindent We also intend to 
   %
    enrich the current application model of \secfog\ so to be able to analyse the security of (probabilistic) information flows among the constituent services, by also defining pre-defined patterns (along the lines of \cite{tsankov2018security}).

%% file: main.bbl
\begin{thebibliography}{70}
\expandafter\ifx\csname natexlab\endcsname\relax\def\natexlab#1{#1}\fi
\providecommand{\bibinfo}[2]{#2}
\ifx\xfnm\relax \def\xfnm[#1]{\unskip,\space#1}\fi
\bibitem[{Fajjari et~al.(2018)Fajjari, Tobagi, and
  Takahashi}]{cloudedgecomputing}
\bibinfo{author}{I.~Fajjari}, \bibinfo{author}{F.~Tobagi},
  \bibinfo{author}{Y.~Takahashi}, \bibinfo{title}{Cloud edge computing in the
  iot}, \bibinfo{year}{2018}.
\bibitem[{Familiar(2015)}]{familiar2015iot}
\bibinfo{author}{B.~Familiar},
\newblock \bibinfo{title}{Iot and microservices},
\newblock in: \bibinfo{booktitle}{Microservices, IoT, and Azure},
  \bibinfo{publisher}{Springer}, \bibinfo{year}{2015}, pp.
  \bibinfo{pages}{133--163}.
\bibitem[{Brogi et~al.(2019)Brogi, Forti, and Ibrahim}]{book}
\bibinfo{author}{A.~Brogi}, \bibinfo{author}{S.~Forti},
  \bibinfo{author}{A.~Ibrahim},
\newblock \bibinfo{title}{Predictive analysis to support fog application
  deployment},
\newblock in: \bibinfo{booktitle}{Fog and Edge Computing: Principles and
  Paradigms, Rajkumar Buyya and Satish N. Srirama (eds.)},
  \bibinfo{publisher}{Wiley}, \bibinfo{year}{2019}.
\bibitem[{Brogi et~al.(2018)Brogi, Forti, Ibrahim, and Rinaldi}]{bonsai}
\bibinfo{author}{A.~Brogi}, \bibinfo{author}{S.~Forti},
  \bibinfo{author}{A.~Ibrahim}, \bibinfo{author}{L.~Rinaldi},
\newblock \bibinfo{title}{Bonsai in the fog: An active learning lab with fog
  computing},
\newblock in: \bibinfo{booktitle}{Fog and Mobile Edge Computing (FMEC), 2018
  Third International Conference on}, \bibinfo{organization}{IEEE}, pp.
  \bibinfo{pages}{79--86}.
\bibitem[{Gupta et~al.(2017)Gupta, Vahid~Dastjerdi, Ghosh, and Buyya}]{004}
\bibinfo{author}{H.~Gupta}, \bibinfo{author}{A.~Vahid~Dastjerdi},
  \bibinfo{author}{S.~K. Ghosh}, \bibinfo{author}{R.~Buyya},
\newblock \bibinfo{title}{{iFogSim: A toolkit for modeling and simulation of
  resource management techniques in the Internet of Things, Edge and Fog
  computing environments}},
\newblock \bibinfo{journal}{Software: Practice and Experience}
  \bibinfo{volume}{47} (\bibinfo{year}{2017}) \bibinfo{pages}{1275--1296}.
\bibitem[{Guerrero et~al.(2018)Guerrero, Lera, and Juiz}]{101}
\bibinfo{author}{C.~Guerrero}, \bibinfo{author}{I.~Lera},
  \bibinfo{author}{C.~Juiz},
\newblock \bibinfo{title}{A lightweight decentralized service placement policy
  for performance optimization in fog computing},
\newblock \bibinfo{journal}{Journal of Ambient Intelligence and Humanized
  Computing}  (\bibinfo{year}{2018}).
\bibitem[{Skarlat et~al.(2017)Skarlat, Nardelli, Schulte, and Dustdar}]{024}
\bibinfo{author}{O.~Skarlat}, \bibinfo{author}{M.~Nardelli},
  \bibinfo{author}{S.~Schulte}, \bibinfo{author}{S.~Dustdar},
\newblock \bibinfo{title}{Towards qos-aware fog service placement},
\newblock in: \bibinfo{booktitle}{2017 IEEE 1st International Conference on Fog
  and Edge Computing (ICFEC)}, pp. \bibinfo{pages}{89--96}.
\bibitem[{Brogi and Forti(2017)}]{011}
\bibinfo{author}{A.~Brogi}, \bibinfo{author}{S.~Forti},
\newblock \bibinfo{title}{{QoS-Aware Deployment of IoT Applications Through the
  Fog}},
\newblock \bibinfo{journal}{IEEE Internet of Things Journal}
  \bibinfo{volume}{4} (\bibinfo{year}{2017}) \bibinfo{pages}{1185--1192}.
\bibitem[{Brogi et~al.(2017)Brogi, Forti, and Ibrahim}]{012}
\bibinfo{author}{A.~Brogi}, \bibinfo{author}{S.~Forti},
  \bibinfo{author}{A.~Ibrahim},
\newblock \bibinfo{title}{How to best deploy your {F}og applications,
  probably},
\newblock in: \bibinfo{editor}{O.~Rana}, \bibinfo{editor}{R.~Buyya},
  \bibinfo{editor}{A.~Anjum} (Eds.), \bibinfo{booktitle}{Proceedings of 1st
  IEEE Int. Conference on Fog and Edge Computing}.
\bibitem[{Forti et~al.(2019)Forti, Ibrahim, and Brogi}]{summersoc}
\bibinfo{author}{S.~Forti}, \bibinfo{author}{A.~Ibrahim},
  \bibinfo{author}{A.~Brogi},
\newblock \bibinfo{title}{Mimicking fogdirector application management},
\newblock \bibinfo{journal}{Software-Intensive Cyber-Physical Systems}
  \bibinfo{volume}{34} (\bibinfo{year}{2019}) \bibinfo{pages}{151--161}.
\bibitem[{{Brogi} et~al.(2019){Brogi}, {Forti}, {Guerrero}, and
  {Lera}}]{howtoplace2019}
\bibinfo{author}{A.~{Brogi}}, \bibinfo{author}{S.~{Forti}},
  \bibinfo{author}{C.~{Guerrero}}, \bibinfo{author}{I.~{Lera}},
  \bibinfo{title}{{How to Place Your Apps in the Fog - State of the Art and
  Open Challenges}}, \bibinfo{year}{2019}.
\bibitem[{Vigan{\`o} and Magazzeni(2018)}]{vigano2018explainable}
\bibinfo{author}{L.~Vigan{\`o}}, \bibinfo{author}{D.~Magazzeni},
\newblock \bibinfo{title}{Explainable security},
\newblock \bibinfo{journal}{arXiv preprint arXiv:1807.04178}
  (\bibinfo{year}{2018}).
\bibitem[{Ni et~al.(2017)Ni, Zhang, Lin, and Shen}]{ni2017securing}
\bibinfo{author}{J.~Ni}, \bibinfo{author}{K.~Zhang}, \bibinfo{author}{X.~Lin},
  \bibinfo{author}{X.~Shen},
\newblock \bibinfo{title}{Securing fog computing for internet of things
  applications: Challenges and solutions},
\newblock \bibinfo{journal}{IEEE Comm. Surveys \& Tutorials}
  (\bibinfo{year}{2017}).
\bibitem[{Ambrosin et~al.(2016)Ambrosin, Conti, Ibrahim, Neven, Sadeghi, and
  Schunter}]{ambrosin2016sana}
\bibinfo{author}{M.~Ambrosin}, \bibinfo{author}{M.~Conti},
  \bibinfo{author}{A.~Ibrahim}, \bibinfo{author}{G.~Neven},
  \bibinfo{author}{A.-R. Sadeghi}, \bibinfo{author}{M.~Schunter},
\newblock \bibinfo{title}{Sana: secure and scalable aggregate network
  attestation},
\newblock in: \bibinfo{booktitle}{Proceedings of the 2016 ACM SIGSAC Conference
  on Computer and Communications Security}, \bibinfo{organization}{ACM}, pp.
  \bibinfo{pages}{731--742}.
\bibitem[{Torkzaban et~al.(2019)Torkzaban, Papagianni, and Baras}]{fmectrust}
\bibinfo{author}{N.~Torkzaban}, \bibinfo{author}{C.~Papagianni},
  \bibinfo{author}{J.~S. Baras},
\newblock \bibinfo{title}{Trust-aware service chain embedding},
\newblock in: \bibinfo{booktitle}{Proceedings of the 6th International
  Conference on Software Defined Systems (SDS)}.
\bibitem[{Garcia~Lopez et~al.(2015)Garcia~Lopez, Montresor, Epema, Datta,
  Higashino, Iamnitchi, Barcellos, Felber, and Riviere}]{garcia}
\bibinfo{author}{P.~Garcia~Lopez}, \bibinfo{author}{A.~Montresor},
  \bibinfo{author}{D.~Epema}, \bibinfo{author}{A.~Datta},
  \bibinfo{author}{T.~Higashino}, \bibinfo{author}{A.~Iamnitchi},
  \bibinfo{author}{M.~Barcellos}, \bibinfo{author}{P.~Felber},
  \bibinfo{author}{E.~Riviere},
\newblock \bibinfo{title}{Edge-centric computing: Vision and challenges},
\newblock \bibinfo{journal}{ACM SIGCOMM Computer Communication Review}
  \bibinfo{volume}{45} (\bibinfo{year}{2015}) \bibinfo{pages}{37--42}.
\bibitem[{De~Raedt et~al.(2007)De~Raedt, Kimmig, and Toivonen}]{problog07}
\bibinfo{author}{L.~De~Raedt}, \bibinfo{author}{A.~Kimmig},
  \bibinfo{author}{H.~Toivonen},
\newblock \bibinfo{title}{Problog: A probabilistic prolog and its application
  in link discovery},
\newblock in: \bibinfo{booktitle}{Proceedings of the 20th International Joint
  Conference on Artifical Intelligence}, pp. \bibinfo{pages}{2468--2473}.
\bibitem[{De~Raedt and Kimmig(2015)}]{problog15}
\bibinfo{author}{L.~De~Raedt}, \bibinfo{author}{A.~Kimmig},
\newblock \bibinfo{title}{Probabilistic (logic) programming concepts},
\newblock \bibinfo{journal}{Machine Learning} \bibinfo{volume}{100}
  (\bibinfo{year}{2015}) \bibinfo{pages}{5--47}.
\bibitem[{Arcangeli et~al.(2015)Arcangeli, Boujbel, and Leriche}]{arcangeli}
\bibinfo{author}{J.-P. Arcangeli}, \bibinfo{author}{R.~Boujbel},
  \bibinfo{author}{S.~Leriche},
\newblock \bibinfo{title}{Automatic deployment of distributed software systems:
  {D}efinitions and state of the art},
\newblock \bibinfo{journal}{Journal of Systems and Software}
  \bibinfo{volume}{103} (\bibinfo{year}{2015}) \bibinfo{pages}{198--218}.
\bibitem[{ope(2019)}]{openfog}
\bibinfo{title}{{O}pen{F}og {C}onsortium},
  \bibinfo{howpublished}{http://www.openfogconsortium.org/},
  \bibinfo{year}{2019}.
\bibitem[{clo(2014)}]{cloudslaguidelines}
\bibinfo{title}{{EU Cloud SLA Standardisation Guidelines}},
  \bibinfo{year}{2014}.
\bibitem[{Mezni et~al.(2018)Mezni, Sellami, and Kouki}]{mezni2018security}
\bibinfo{author}{H.~Mezni}, \bibinfo{author}{M.~Sellami},
  \bibinfo{author}{J.~Kouki},
\newblock \bibinfo{title}{{Security-aware SaaS placement using swarm
  intelligence}},
\newblock \bibinfo{journal}{Journal of Software: Evolution and Process}
  (\bibinfo{year}{2018}).
\bibitem[{Mukherjee et~al.(2017)Mukherjee, Matam, Shu, Maglaras, Ferrag,
  Choudhury, and Kumar}]{mukherjee2017security}
\bibinfo{author}{M.~Mukherjee}, \bibinfo{author}{R.~Matam},
  \bibinfo{author}{L.~Shu}, \bibinfo{author}{L.~Maglaras},
  \bibinfo{author}{M.~A. Ferrag}, \bibinfo{author}{N.~Choudhury},
  \bibinfo{author}{V.~Kumar},
\newblock \bibinfo{title}{Security and privacy in fog computing: Challenges},
\newblock \bibinfo{journal}{IEEE Access} \bibinfo{volume}{5}
  (\bibinfo{year}{2017}) \bibinfo{pages}{19293--19304}.
\bibitem[{Rodr{\'\i}guez and Egenhofer(2003)}]{mediation}
\bibinfo{author}{M.~A. Rodr{\'\i}guez}, \bibinfo{author}{M.~J. Egenhofer},
\newblock \bibinfo{title}{Determining semantic similarity among entity classes
  from different ontologies},
\newblock \bibinfo{journal}{IEEE transactions on knowledge and data
  engineering} \bibinfo{volume}{15} (\bibinfo{year}{2003})
  \bibinfo{pages}{442--456}.
\bibitem[{Theodorakopoulos and Baras(2006)}]{trust06}
\bibinfo{author}{G.~Theodorakopoulos}, \bibinfo{author}{J.~S. Baras},
\newblock \bibinfo{title}{On trust models and trust evaluation metrics for
  ad-hoc networks},
\newblock \bibinfo{journal}{IEEE Journal on selected areas in Communications}
  \bibinfo{volume}{24} (\bibinfo{year}{2006}) \bibinfo{pages}{318--328}.
\bibitem[{Tang et~al.(2017)Tang, Dai, Liu, and Chen}]{TDLC17}
\bibinfo{author}{M.~Tang}, \bibinfo{author}{X.~Dai}, \bibinfo{author}{J.~Liu},
  \bibinfo{author}{J.~Chen},
\newblock \bibinfo{title}{Towards a trust evaluation middleware for cloud
  service selection},
\newblock \bibinfo{journal}{Future Generation Computer Systems}
  \bibinfo{volume}{74} (\bibinfo{year}{2017}) \bibinfo{pages}{302--312}.
\bibitem[{Brogi et~al.(2019)Brogi, Forti, and Ibrahim}]{ccisbrogiforti}
\bibinfo{author}{A.~Brogi}, \bibinfo{author}{S.~Forti},
  \bibinfo{author}{A.~Ibrahim},
\newblock \bibinfo{title}{{Optimising QoS-assurance, Resource Usage and Cost of
  Fog Application Deployments}},
\newblock in: \bibinfo{booktitle}{Cloud Computing and Services Science Selected
  Papers, Communications in Computer and Information Science},
  \bibinfo{publisher}{Springer}, \bibinfo{year}{2019}. \bibinfo{note}{In
  Press}.
\bibitem[{Newman(2015)}]{microservices}
\bibinfo{author}{S.~Newman}, \bibinfo{title}{Building microservices: designing
  fine-grained systems}, \bibinfo{publisher}{" O'Reilly Media, Inc."},
  \bibinfo{year}{2015}.
\bibitem[{Yan et~al.(2007)}]{yan2007trust}
\bibinfo{author}{Z.~Yan}, et~al., \bibinfo{title}{Trust management for mobile
  computing platforms}, \bibinfo{publisher}{Helsinki University of Technology},
  \bibinfo{year}{2007}.
\bibitem[{Bistarelli et~al.(2008)Bistarelli, Martinelli, and
  Santini}]{bistarelli2008weighted}
\bibinfo{author}{S.~Bistarelli}, \bibinfo{author}{F.~Martinelli},
  \bibinfo{author}{F.~Santini},
\newblock \bibinfo{title}{Weighted datalog and levels of trust},
\newblock in: \bibinfo{booktitle}{Availability, Reliability and Security, 2008.
  ARES 08. Third International Conference on}, \bibinfo{organization}{IEEE},
  pp. \bibinfo{pages}{1128--1134}.
\bibitem[{Bistarelli et~al.(2010)Bistarelli, Foley, O'Sullivan, and
  Santini}]{bistarelli2010semiring}
\bibinfo{author}{S.~Bistarelli}, \bibinfo{author}{S.~N. Foley},
  \bibinfo{author}{B.~O'Sullivan}, \bibinfo{author}{F.~Santini},
\newblock \bibinfo{title}{Semiring-based frameworks for trust propagation in
  small-world networks and coalition formation criteria},
\newblock \bibinfo{journal}{Security and Communication Networks}
  \bibinfo{volume}{3} (\bibinfo{year}{2010}) \bibinfo{pages}{595--610}.
\bibitem[{Gao et~al.(2016)Gao, Miao, Baras, and Golbeck}]{gao2016star}
\bibinfo{author}{P.~Gao}, \bibinfo{author}{H.~Miao}, \bibinfo{author}{J.~S.
  Baras}, \bibinfo{author}{J.~Golbeck},
\newblock \bibinfo{title}{Star: Semiring trust inference for trust-aware social
  recommenders},
\newblock in: \bibinfo{booktitle}{Proceedings of the 10th ACM Conference on
  Recommender Systems}, \bibinfo{organization}{ACM}, pp.
  \bibinfo{pages}{301--308}.
\bibitem[{Kimmig et~al.(2011)Kimmig, Van~den Broeck, and
  De~Raedt}]{kimmig2011algebraic}
\bibinfo{author}{A.~Kimmig}, \bibinfo{author}{G.~Van~den Broeck},
  \bibinfo{author}{L.~De~Raedt},
\newblock \bibinfo{title}{An algebraic prolog for reasoning about possible
  worlds},
\newblock in: \bibinfo{booktitle}{Twenty-Fifth AAAI Conference on Artificial
  Intelligence}.
\bibitem[{Kaur et~al.(2017)Kaur, Singh, Singh et~al.}]{kaur2017taxonomy}
\bibinfo{author}{A.~Kaur}, \bibinfo{author}{M.~Singh},
  \bibinfo{author}{P.~Singh}, et~al.,
\newblock \bibinfo{title}{A taxonomy, survey on placement of virtual machines
  in cloud},
\newblock in: \bibinfo{booktitle}{2017 International Conference on Energy,
  Communication, Data Analytics and Soft Computing (ICECDS)},
  \bibinfo{organization}{IEEE}, pp. \bibinfo{pages}{2054--2058}.
\bibitem[{Schoenen et~al.(2017)Schoenen, Mann, and Metzger}]{zoltan}
\bibinfo{author}{S.~Schoenen}, \bibinfo{author}{Z.~{\'A}. Mann},
  \bibinfo{author}{A.~Metzger},
\newblock \bibinfo{title}{Using risk patterns to identify violations of data
  protection policies in cloud systems},
\newblock in: \bibinfo{booktitle}{International Conference on Service-Oriented
  Computing}, \bibinfo{organization}{Springer}, pp. \bibinfo{pages}{296--307}.
\bibitem[{Nacer et~al.(2016)Nacer, Goettelmann, Youcef, Tari, and
  Godart}]{nacer2016obfuscating}
\bibinfo{author}{A.~A. Nacer}, \bibinfo{author}{E.~Goettelmann},
  \bibinfo{author}{S.~Youcef}, \bibinfo{author}{A.~Tari},
  \bibinfo{author}{C.~Godart},
\newblock \bibinfo{title}{Obfuscating a business process by splitting its logic
  with fake fragments for securing a multi-cloud deployment},
\newblock in: \bibinfo{booktitle}{Services (SERVICES), 2016 IEEE World Congress
  on}, \bibinfo{organization}{IEEE}, pp. \bibinfo{pages}{18--25}.
\bibitem[{Goettelmann et~al.(2014)Goettelmann, Dahman, Gateau, Dubois, and
  Godart}]{goettelmann2014security}
\bibinfo{author}{E.~Goettelmann}, \bibinfo{author}{K.~Dahman},
  \bibinfo{author}{B.~Gateau}, \bibinfo{author}{E.~Dubois},
  \bibinfo{author}{C.~Godart},
\newblock \bibinfo{title}{A security risk assessment model for business process
  deployment in the cloud},
\newblock in: \bibinfo{booktitle}{Services Computing (SCC), 2014 IEEE
  International Conference on}, \bibinfo{organization}{IEEE}, pp.
  \bibinfo{pages}{307--314}.
\bibitem[{Wen et~al.(2017)Wen, Ca{\l}a, Watson, and Romanovsky}]{wen2017cost}
\bibinfo{author}{Z.~Wen}, \bibinfo{author}{J.~Ca{\l}a},
  \bibinfo{author}{P.~Watson}, \bibinfo{author}{A.~Romanovsky},
\newblock \bibinfo{title}{Cost effective, reliable and secure workflow
  deployment over federated clouds},
\newblock \bibinfo{journal}{IEEE Transactions on Services Computing}
  \bibinfo{volume}{10} (\bibinfo{year}{2017}) \bibinfo{pages}{929--941}.
\bibitem[{Luna et~al.(2017)Luna, Taha, Trapero, and Suri}]{luna}
\bibinfo{author}{J.~Luna}, \bibinfo{author}{A.~Taha},
  \bibinfo{author}{R.~Trapero}, \bibinfo{author}{N.~Suri},
\newblock \bibinfo{title}{Quantitative reasoning about cloud security using
  service level agreements},
\newblock \bibinfo{journal}{IEEE Transactions on Cloud Computing}
  \bibinfo{volume}{5} (\bibinfo{year}{2017}) \bibinfo{pages}{457--471}.
\bibitem[{Varshney and Simmhan(2017)}]{icfecreview}
\bibinfo{author}{P.~Varshney}, \bibinfo{author}{Y.~Simmhan},
\newblock \bibinfo{title}{Demystifying {F}og {C}omputing: {C}haracterizing
  {A}rchitectures, {A}pplications and {A}bstractions},
\newblock in: \bibinfo{booktitle}{2017 IEEE 1st International Conference on Fog
  and Edge Computing (ICFEC)}, pp. \bibinfo{pages}{115--124}.
\bibitem[{Wen et~al.(2017)Wen, Yang, Garraghan, Lin, Xu, and
  Rovatsos}]{recentreview}
\bibinfo{author}{Z.~Wen}, \bibinfo{author}{R.~Yang},
  \bibinfo{author}{P.~Garraghan}, \bibinfo{author}{T.~Lin},
  \bibinfo{author}{J.~Xu}, \bibinfo{author}{M.~Rovatsos},
\newblock \bibinfo{title}{Fog {O}rchestration for {I}nternet of {T}hings
  {S}ervices},
\newblock \bibinfo{journal}{IEEE Internet Computing} \bibinfo{volume}{21}
  (\bibinfo{year}{2017}) \bibinfo{pages}{16--24}.
\bibitem[{Mahmud et~al.(2018)Mahmud, Ramamohanarao, and Buyya}]{021}
\bibinfo{author}{R.~Mahmud}, \bibinfo{author}{K.~Ramamohanarao},
  \bibinfo{author}{R.~Buyya},
\newblock \bibinfo{title}{Latency-aware application module management for fog
  computing environments},
\newblock \bibinfo{journal}{ACM Transactions on Internet Technology (TOIT)}
  (\bibinfo{year}{2018}).
\bibitem[{Wang et~al.(2017)Wang, Zafer, and Leung}]{032}
\bibinfo{author}{S.~Wang}, \bibinfo{author}{M.~Zafer}, \bibinfo{author}{K.~K.
  Leung},
\newblock \bibinfo{title}{Online placement of multi-component applications in
  edge computing environments},
\newblock \bibinfo{journal}{IEEE Access} \bibinfo{volume}{5}
  (\bibinfo{year}{2017}) \bibinfo{pages}{2514--2533}.
\bibitem[{Brogi et~al.(2018)Brogi, Forti, and Ibrahim}]{closer18}
\bibinfo{author}{A.~Brogi}, \bibinfo{author}{S.~Forti},
  \bibinfo{author}{A.~Ibrahim},
\newblock \bibinfo{title}{Deploying fog applications: How much does it cost, by
  the way?},
\newblock in: \bibinfo{booktitle}{Proceedings of the 8th International
  Conference on Cloud Computing and Services Science},
  \bibinfo{publisher}{SciTePress}, \bibinfo{year}{2018}, pp.
  \bibinfo{pages}{68--77}.
\bibitem[{Hong et~al.(2016)Hong, Tsai, and Hsu}]{062}
\bibinfo{author}{H.~J. Hong}, \bibinfo{author}{P.~H. Tsai},
  \bibinfo{author}{C.~H. Hsu},
\newblock \bibinfo{title}{Dynamic module deployment in a fog computing
  platform},
\newblock in: \bibinfo{booktitle}{2016 18th Asia-Pacific Network Operations and
  Management Symposium (APNOMS)}, pp. \bibinfo{pages}{1--6}.
\bibitem[{Taneja and Davy(2017)}]{027}
\bibinfo{author}{M.~Taneja}, \bibinfo{author}{A.~Davy},
\newblock \bibinfo{title}{Resource aware placement of iot application modules
  in fog-cloud computing paradigm},
\newblock in: \bibinfo{booktitle}{2017 IFIP/IEEE Symposium on Integrated
  Network and Service Management (IM)}, pp. \bibinfo{pages}{1222--1228}.
\bibitem[{Hamid Reza~Arkian(2017)}]{001}
\bibinfo{author}{A.~P. Hamid Reza~Arkian, Abolfazl~Diyanat},
\newblock \bibinfo{title}{Mist: Fog-based data analytics scheme with
  cost-efficient resource provisioning for {I}o{T} crowdsensing applications},
\newblock \bibinfo{journal}{Journal of Network and Computer Applications}
  \bibinfo{volume}{82} (\bibinfo{year}{2017}) \bibinfo{pages}{152 -- 165}.
\bibitem[{Yang et~al.(2016)Yang, Cao, Liang, and Han}]{034}
\bibinfo{author}{L.~Yang}, \bibinfo{author}{J.~Cao},
  \bibinfo{author}{G.~Liang}, \bibinfo{author}{X.~Han},
\newblock \bibinfo{title}{Cost aware service placement and load dispatching in
  mobile cloud systems},
\newblock \bibinfo{journal}{IEEE Transactions on Computers}
  \bibinfo{volume}{65} (\bibinfo{year}{2016}) \bibinfo{pages}{1440--1452}.
\bibitem[{Gu et~al.(2017)Gu, Zeng, Guo, Barnawi, and Xiang}]{017}
\bibinfo{author}{L.~Gu}, \bibinfo{author}{D.~Zeng}, \bibinfo{author}{S.~Guo},
  \bibinfo{author}{A.~Barnawi}, \bibinfo{author}{Y.~Xiang},
\newblock \bibinfo{title}{Cost efficient resource management in fog computing
  supported medical cyber-physical system},
\newblock \bibinfo{journal}{IEEE Transactions on Emerging Topics in Computing}
  \bibinfo{volume}{5} (\bibinfo{year}{2017}) \bibinfo{pages}{108--119}.
\bibitem[{Zeng et~al.(2016)Zeng, Gu, Guo, Cheng, and Yu}]{095}
\bibinfo{author}{D.~Zeng}, \bibinfo{author}{L.~Gu}, \bibinfo{author}{S.~Guo},
  \bibinfo{author}{Z.~Cheng}, \bibinfo{author}{S.~Yu},
\newblock \bibinfo{title}{Joint optimization of task scheduling and image
  placement in fog computing supported software-defined embedded system},
\newblock \bibinfo{journal}{IEEE Transactions on Computers}
  \bibinfo{volume}{65} (\bibinfo{year}{2016}) \bibinfo{pages}{3702--3712}.
\bibitem[{Souza et~al.(2016)Souza, Ramírez, Masip-Bruin, Marín-Tordera, Ren,
  and Tashakor}]{061}
\bibinfo{author}{V.~B.~C. Souza}, \bibinfo{author}{W.~Ramírez},
  \bibinfo{author}{X.~Masip-Bruin}, \bibinfo{author}{E.~Marín-Tordera},
  \bibinfo{author}{G.~Ren}, \bibinfo{author}{G.~Tashakor},
\newblock \bibinfo{title}{Handling service allocation in combined fog-cloud
  scenarios},
\newblock in: \bibinfo{booktitle}{2016 IEEE International Conference on
  Communications (ICC)}, pp. \bibinfo{pages}{1--5}.
\bibitem[{Barcelo et~al.(2016)Barcelo, Correa, Llorca, Tulino, Vicario, and
  Morell}]{002}
\bibinfo{author}{M.~Barcelo}, \bibinfo{author}{A.~Correa},
  \bibinfo{author}{J.~Llorca}, \bibinfo{author}{A.~M. Tulino},
  \bibinfo{author}{J.~L. Vicario}, \bibinfo{author}{A.~Morell},
\newblock \bibinfo{title}{{I}o{T}-cloud service optimization in next generation
  smart environments},
\newblock \bibinfo{journal}{IEEE Journal on Selected Areas in Communications}
  \bibinfo{volume}{34} (\bibinfo{year}{2016}) \bibinfo{pages}{4077--4090}.
\bibitem[{Huang et~al.(2014)Huang, Lin, Yu, and Hsu}]{019}
\bibinfo{author}{Z.~Huang}, \bibinfo{author}{K.-J. Lin}, \bibinfo{author}{S.-Y.
  Yu}, \bibinfo{author}{J.~Y.-j. Hsu},
\newblock \bibinfo{title}{Co-locating services in {I}o{T} systems to minimize
  the communication energy cost},
\newblock \bibinfo{journal}{Journal of Innovation in Digital Ecosystems}
  \bibinfo{volume}{1} (\bibinfo{year}{2014}) \bibinfo{pages}{47--57}.
\bibitem[{Mahmud et~al.(2018)Mahmud, Srirama, Ramamohanarao, and Buyya}]{107}
\bibinfo{author}{R.~Mahmud}, \bibinfo{author}{S.~N. Srirama},
  \bibinfo{author}{K.~Ramamohanarao}, \bibinfo{author}{R.~Buyya},
\newblock \bibinfo{title}{Quality of experience (qoe)-aware placement of
  applications in fog computing environments},
\newblock \bibinfo{journal}{Journal of Parallel and Distributed Computing}
  (\bibinfo{year}{2018}).
\bibitem[{Venticinque and Amato(2018)}]{030}
\bibinfo{author}{S.~Venticinque}, \bibinfo{author}{A.~Amato},
\newblock \bibinfo{title}{A methodology for deployment of iot application in
  fog},
\newblock \bibinfo{journal}{Journal of Ambient Intelligence and Humanized
  Computing}  (\bibinfo{year}{2018}).
\bibitem[{Cardellini et~al.(2016)Cardellini, Grassi, Lo~Presti, and
  Nardelli}]{038}
\bibinfo{author}{V.~Cardellini}, \bibinfo{author}{V.~Grassi},
  \bibinfo{author}{F.~Lo~Presti}, \bibinfo{author}{M.~Nardelli},
\newblock \bibinfo{title}{Optimal operator placement for distributed stream
  processing applications},
\newblock in: \bibinfo{booktitle}{Proceedings of the 10th ACM International
  Conference on Distributed and Event-based Systems}, DEBS '16,
  \bibinfo{publisher}{ACM}, \bibinfo{address}{New York, NY, USA},
  \bibinfo{year}{2016}, pp. \bibinfo{pages}{69--80}.
\bibitem[{Rahbari and Nickray(2017)}]{006}
\bibinfo{author}{D.~Rahbari}, \bibinfo{author}{M.~Nickray},
\newblock \bibinfo{title}{Scheduling of fog networks with optimized knapsack by
  symbiotic organisms search},
\newblock in: \bibinfo{booktitle}{2017 21st Conference of Open Innovations
  Association (FRUCT)}, pp. \bibinfo{pages}{278--283}.
\bibitem[{{Brogi} et~al.(????){Brogi}, {Forti}, {Guerrero}, and
  {Lera}}]{geneticalgorithms2019}
\bibinfo{author}{A.~{Brogi}}, \bibinfo{author}{S.~{Forti}},
  \bibinfo{author}{C.~{Guerrero}}, \bibinfo{author}{I.~{Lera}},
\newblock \bibinfo{title}{{Meet Genetic Algorithms in Monte Carlo: Optimised
  Placement of Multi-Service Applications in the Fog}},
\newblock in: \bibinfo{booktitle}{{Proceedings of the 3rd IEEE International
  Conference on Edge Computing (EDGE 2019)}}, pp. \bibinfo{pages}{13--17}.
\bibitem[{Tang et~al.(2018)Tang, Zhou, Zhang, Jia, and Zhao}]{008}
\bibinfo{author}{Z.~Tang}, \bibinfo{author}{X.~Zhou},
  \bibinfo{author}{F.~Zhang}, \bibinfo{author}{W.~Jia},
  \bibinfo{author}{W.~Zhao},
\newblock \bibinfo{title}{Migration modeling and learning algorithms for
  containers in fog computing},
\newblock \bibinfo{journal}{IEEE Transactions on Services Computing}
  (\bibinfo{year}{2018}).
\bibitem[{Ardagna et~al.(2007)Ardagna, Damiani, De~Capitani~di Vimercati,
  Foresti, and Samarati}]{trustmanagementardagna}
\bibinfo{author}{C.~A. Ardagna}, \bibinfo{author}{E.~Damiani},
  \bibinfo{author}{S.~De~Capitani~di Vimercati}, \bibinfo{author}{S.~Foresti},
  \bibinfo{author}{P.~Samarati}, \bibinfo{title}{Trust Management},
  \bibinfo{publisher}{Springer Berlin Heidelberg}, \bibinfo{address}{Berlin,
  Heidelberg}, pp. \bibinfo{pages}{103--117}.
\bibitem[{Li et~al.(2002)Li, Mitchell, and Winsborough}]{li2002design}
\bibinfo{author}{N.~Li}, \bibinfo{author}{J.~C. Mitchell},
  \bibinfo{author}{W.~H. Winsborough},
\newblock \bibinfo{title}{Design of a role-based trust-management framework},
\newblock in: \bibinfo{booktitle}{Security and Privacy, 2002. Proceedings. 2002
  IEEE Symposium on}, \bibinfo{organization}{IEEE}, pp.
  \bibinfo{pages}{114--130}.
\bibitem[{Ziegler and Lausen(2005)}]{ziegler2005propagation}
\bibinfo{author}{C.-N. Ziegler}, \bibinfo{author}{G.~Lausen},
\newblock \bibinfo{title}{Propagation models for trust and distrust in social
  networks},
\newblock \bibinfo{journal}{Information Systems Frontiers} \bibinfo{volume}{7}
  (\bibinfo{year}{2005}) \bibinfo{pages}{337--358}.
\bibitem[{Twigg and Dimmock(2003)}]{twigg2003attack}
\bibinfo{author}{A.~Twigg}, \bibinfo{author}{N.~Dimmock},
\newblock \bibinfo{title}{Attack-resistance of computational trust models},
\newblock in: \bibinfo{booktitle}{Enabling Technologies: Infrastructure for
  Collaborative Enterprises, 2003. WET ICE 2003. Proceedings. Twelfth IEEE
  International Workshops on}, \bibinfo{organization}{IEEE}, pp.
  \bibinfo{pages}{275--280}.
\bibitem[{Omar et~al.(2012)Omar, Challal, and
  Bouabdallah}]{omar2012certification}
\bibinfo{author}{M.~Omar}, \bibinfo{author}{Y.~Challal},
  \bibinfo{author}{A.~Bouabdallah},
\newblock \bibinfo{title}{Certification-based trust models in mobile ad hoc
  networks: A survey and taxonomy},
\newblock \bibinfo{journal}{Journal of Network and Computer Applications}
  \bibinfo{volume}{35} (\bibinfo{year}{2012}) \bibinfo{pages}{268--286}.
\bibitem[{Anisetti et~al.(2014)Anisetti, Ardagna, and
  Damiani}]{anisetti2014certification}
\bibinfo{author}{M.~Anisetti}, \bibinfo{author}{C.~A. Ardagna},
  \bibinfo{author}{E.~Damiani},
\newblock \bibinfo{title}{A certification-based trust model for autonomic cloud
  computing systems},
\newblock in: \bibinfo{booktitle}{Cloud and Autonomic Computing (ICCAC), 2014
  International Conference on}, \bibinfo{organization}{IEEE}, pp.
  \bibinfo{pages}{212--219}.
\bibitem[{Anisetti et~al.(2017)Anisetti, Ardagna, Damiani, and
  Gaudenzi}]{anisetti2017semi}
\bibinfo{author}{M.~Anisetti}, \bibinfo{author}{C.~Ardagna},
  \bibinfo{author}{E.~Damiani}, \bibinfo{author}{F.~Gaudenzi},
\newblock \bibinfo{title}{A semi-automatic and trustworthy scheme for
  continuous cloud service certification},
\newblock \bibinfo{journal}{IEEE Transactions on Services Computing}
  (\bibinfo{year}{2017}).
\bibitem[{Ding et~al.(2017)Ding, Wang, Wu, and Olson}]{DWWO17}
\bibinfo{author}{S.~Ding}, \bibinfo{author}{Z.~Wang}, \bibinfo{author}{D.~Wu},
  \bibinfo{author}{D.~L. Olson},
\newblock \bibinfo{title}{Utilizing customer satisfaction in ranking prediction
  for personalized cloud service selection},
\newblock \bibinfo{journal}{Decision Support Systems} \bibinfo{volume}{93}
  (\bibinfo{year}{2017}) \bibinfo{pages}{1--10}.
\bibitem[{Qu et~al.(2015)Qu, Wang, Orgun, Liu, Liu, and Bouguettaya}]{QWOLLB15}
\bibinfo{author}{L.~Qu}, \bibinfo{author}{Y.~Wang}, \bibinfo{author}{M.~A.
  Orgun}, \bibinfo{author}{L.~Liu}, \bibinfo{author}{H.~Liu},
  \bibinfo{author}{A.~Bouguettaya},
\newblock \bibinfo{title}{Cccloud: Context-aware and credible cloud service
  selection based on subjective assessment and objective assessment},
\newblock \bibinfo{journal}{IEEE Transactions on Services Computing}
  \bibinfo{volume}{8} (\bibinfo{year}{2015}) \bibinfo{pages}{369--383}.
\bibitem[{{De Capitani Di Vimercati} et~al.(2017){De Capitani Di Vimercati},
  {Foresti}, {Livraga}, {Piuri}, and {Samarati}}]{reviewersuggestion}
\bibinfo{author}{S.~{De Capitani Di Vimercati}},
  \bibinfo{author}{S.~{Foresti}}, \bibinfo{author}{G.~{Livraga}},
  \bibinfo{author}{V.~{Piuri}}, \bibinfo{author}{P.~{Samarati}},
\newblock \bibinfo{title}{Supporting user requirements and preferences in cloud
  plan selection},
\newblock \bibinfo{journal}{IEEE Transactions on Services Computing}
  (\bibinfo{year}{2017}). \bibinfo{note}{In press}.
\bibitem[{Tsankov(2018)}]{tsankov2018security}
\bibinfo{author}{P.~Tsankov},
\newblock \bibinfo{title}{{Security Analysis of Smart Contracts in Datalog}},
\newblock in: \bibinfo{booktitle}{International Symposium on Leveraging
  Applications of Formal Methods}, \bibinfo{organization}{Springer}, pp.
  \bibinfo{pages}{316--322}.

\end{thebibliography}
